\documentclass{aa}

\usepackage{graphicx}
\usepackage{txfonts}
\usepackage{natbib}
\usepackage{amsmath}
\usepackage{bbold}
\usepackage{tikz}
\usepackage{multirow}
\usepackage{xspace}
\usepackage[colorlinks=true, linkcolor=blue, citecolor=blue, filecolor=blue, urlcolor=blue]{hyperref}
\usepackage{placeins}

\def\OGP{{\rm OGP}\xspace}
\def\ILP{{\rm ILP}\xspace}
\def\OGPs{{\rm OGPs}\xspace}
\def\ILPs{{\rm ILPs}\xspace}

\def\targetI{HD~23079}
\def\mtargetI{$8.3~m_\oplus$}
\def\ptargetI{5.75~d}

\def\targetII{HD~196067}
\def\mtargetII{$10.4~m_\oplus$}
\def\ptargetII{4.6~d}

\def\targetIII{HD~86226}
\def\mtargetIII{$7.5~m_\oplus$}
\def\ptargetIII{3.98~d}

\def\targets{\targetI, \targetII, and \targetIII}

\newcommand{\ms}{\ensuremath{\mathrm{m} \, \mathrm{s}^{-1}}\xspace}

\graphicspath{{./figures/}}

\bibpunct{(}{)}{;}{a}{}{,}

\begin{document}
\title{Architecture of planetary systems with and without outer giant planets}
\subtitle{I. Inner planet detections around \object{HD~23079},~\object{HD~196067}, and \object{HD~86226}}
\titlerunning{Architecture of planetary systems with and without giant planets I. Inner planet detections}

\author{J.-B.~Delisle\inst{1, \orcidlink{0000-0001-5844-9888}}
  \and J.~P.~Faria\inst{1, \orcidlink{0000-0002-6728-244X}}
  \and D.~Ségransan\inst{1, \orcidlink{0000-0003-2355-8034}}
  \and E.~Fontanet\inst{1, \orcidlink{0000-0002-0215-4551}}
  \and W.~Ceva\inst{1, \orcidlink{0009-0009-7691-6027}}
  \and D.~Barbato\inst{2, \orcidlink{0009-0005-9862-2549}}
  \and S.~G.~Sousa\inst{11,12, \orcidlink{0000-0001-9047-2965}}
  \and N.~Unger\inst{1, \orcidlink{0000-0003-3993-7127}}
  \and A.~Leleu\inst{1, \orcidlink{0000-0003-2051-7974}}
  \and F.~Bouchy\inst{1, \orcidlink{0000-0002-7613-393X}}
  \and M.~Cretignier\inst{3, \orcidlink{0000-0002-2207-0750}}
  \and R.~F.~D\'iaz\inst{4,5, \orcidlink{0000-0001-9289-5160}}
  \and X.~Dumusque\inst{1, \orcidlink{0000-0002-9332-2011}}
  \and Y.~G.~C.~Frensch\inst{1, \orcidlink{0000-0003-4009-0330}}
  \and N.~C.~Hara\inst{6, \orcidlink{0000-0001-9232-3314}}
  \and G.~Laughlin\inst{7, \orcidlink{0000-0002-3253-2621}}
  \and G.~Lo~Curto\inst{8, \orcidlink{0000-0002-1158-9354}}
  \and C.~Lovis\inst{1, \orcidlink{0000-0001-7120-5837}}
  \and M.~Marmier\inst{1, \orcidlink{0000-0001-5630-1396}}
  \and M.~Mayor\inst{1, \orcidlink{0000-0002-9352-5935}}
  \and L.~Mignon\inst{1,9, \orcidlink{0000-0002-5407-3905}}
  \and C.~Mordasini\inst{10, \orcidlink{0000-0002-1013-2811}}
  \and F.~Pepe\inst{1, \orcidlink{0000-0002-9815-773X}}
  \and N.~C.~Santos\inst{11,12, \orcidlink{0000-0003-4422-2919}}
  \and S.~Udry\inst{1, \orcidlink{0000-0001-7576-6236}}
}

\institute{Département d'astronomie, Université de Genève, chemin Pegasi 51, 1290 Versoix, Switzerland\\ \email{jean-baptiste.delisle@unige.ch}
  \and INAF – Osservatorio Astronomico di Padova, Vicolo dell’Osservatorio 5, 35122 Padova, Italy
  \and Department of Physics, University of Oxford, OX13RH Oxford, UK
  \and Instituto Tecnol\'ogico de Buenos Aires (ITBA), Iguaz\'u 341, Buenos Aires, CABA C1437, Argentina
  \and Instituto de Ciencias F\'isicas (CONICET / ECyT-UNSAM), Campus Miguelete, 25 de Mayo y Francia, (1650) Buenos Aires, Argentina.
  \and Aix Marseille Université, CNRS, CNES, LAM, Marseille, France
  \and Department of Astronomy, Yale University, 52 Hillhouse Avenue, New Haven, CT 06511, USA
  \and European Southern Observatory, Av. Alonso de Cordova 3107, Vitacura, Casilla 19001, Santiago de Chile, Chile
  \and Univ. Grenoble Alpes, CNRS, IPAG, 38000 Grenoble, France
  \and Division of Space Research and Planetary Sciences, Physics Institute, University of Bern, Gesellschaftsstrasse 6, CH-3012 Bern, Switzerland
  \and Instituto de Astrof\'isica e Ci\^encias do Espa\c{c}o, Universidade do Porto, CAUP, Rua das Estrelas, 4150-762 Porto, Portugal
  \and Departamento de F\'isica e Astronomia, Faculdade de Ci\^encias, Universidade do Porto, Rua do Campo Alegre, 4169-007 Porto, Portugal
}

\date{\today}

\abstract{
  Understanding the link between outer giant planets (\OGPs)
  and inner light planets (\ILPs) is key to understanding
  planetary system formation and architecture.
  The correlation between these two populations of planets is debated both theoretically
  -- different formation models predict either a correlation or an anticorrelation --
  and observationally.
  Several recent attempts to constrain this correlation have yielded contradictory results,
  due to small-number statistics and heterogeneous samples.
  We present an ongoing long-term observational effort with CORALIE, HARPS, and ESPRESSO
  to probe the \ILP occurrence in systems with and without \OGP.
  In this first article of a series,
  we discuss how, from the design to the observations,
  we ensured the homogeneity of the samples,
  both in terms of stellar properties and observing strategy.
  We also present the first three detections of \ILPs
  in our \OGP host sample.
  We find a \mtargetI{} planet at \ptargetI{} around \targetI{},
  a \mtargetII{} planet at \ptargetII{} around \targetII{},
  and we confirm the \mtargetIII{} planet at \ptargetIII{} around \targetIII{}.
  While a rigorous statistical analysis of our samples will be performed in subsequent studies,
  the relatively low number of detections in our sample seems to contradict previous studies that found a strong \OGP-\ILP correlation.
}


\maketitle

\section{Introduction}
\label{sec:introduction}

Characterizing the global architecture of planetary systems is crucial for
understanding their formation history and dynamical evolution. Although more than
5000 exoplanets have already been detected\footnote{(e.g., \url{https://exoplanet.eu})},
it remains extremely challenging to
fully characterize both the inner and outer regions of planetary systems due to
the disparate planetary orbital periods, ranging from hours to decades, and
the detection limits and biases of the different observing methods (transit,
radial velocities, imaging, etc.). Thus, the coupling between the inner and
outer parts of planetary systems during planet formation and
subsequent evolution remains poorly constrained by observations. In particular,
the question of whether the presence of outer giant planets (hereafter \OGPs)
enhances or inhibits the formation of inner light planets (hereafter \ILPs) remains open, both theoretically and observationally.

The main planet formation models have different theoretical
predictions. In the inward migration theory, in which \ILPs are initially formed in
the outer part of the protoplanetary disk and migrate to the inner part due to
disk-planet interactions, the presence of an \OGP is proposed to hinder inward migration
\citep[e.g.,][]{izidoro_2015_giant}.
Similarly, in the pebble accretion theory,
\ILPs form in the inner part of the disk owing to a flux of pebbles drifting
inward from the outer region.
The presence of an \OGP would, in that case,
inhibit the pebble flux and thus the formation of \ILPs
\citep[e.g.,][]{rice_2006_dust,ormel_2017_formation,lambrechts_2019_formation}.
Thus, these formation models predict an anticorrelation between \ILP
and \OGP populations. However, in the in situ formation model, orbital
migration is assumed to play a limited role, and the protoplanetary disk is initially
assumed to be sufficiently massive for \ILPs to form in the inner region
by accretion of solids and gas in their vicinity. Such a massive disk also
tends to form gas giant planets in its outer region
\citep[e.g.,][]{chiang_2013_minimummass,schlecker_2021_generation}. Thus, a
correlation between \ILPs and \OGPs would be expected in the in situ formation
model.

Several recent studies have attempted to constrain this correlation from
observations, leading to contradictory results. Most of these studies rely on
comparing the occurrence rates of \OGPs in blind radial velocity (RV)
surveys and in RV follow-up of transiting \ILP hosts. The first attempts
\citep{zhu_2018_super,bryan_2019_excess} find an enhanced probability (by a
factor of three, albeit with large uncertainty)
of \OGP around transiting \ILP hosts compared to the general
population (\OGPs in blind RV surveys).
From this, they conclude that stars hosting
an \OGP have about a 90\% chance of also hosting \ILPs.

At the same time, a
HARPS (High Accuracy Radial velocity Planet Searcher)
search for inner planets in 20 \OGP hosts\citep{barbato_2018_exploring} yielded no \ILP detection.
However, the relatively small sample of stars and
the low number of HARPS measurements per star (27 on average) did not allow the
authors to draw firm conclusions.
A statistical analysis of RV archival data
from the California Legacy Survey \citep[CLS,][]{rosenthal_2021_california},
shows a nonsignificant (within 1-$\sigma$) correlation between \OGPs and \ILPs
\citep{rosenthal_2022_california}.
These approaches \citep{barbato_2018_exploring,rosenthal_2021_california},
as well as the one proposed here,
are direct measurements of $P(\ILP|\OGP)$, in contrast to the indirect method
proposed by \citet{zhu_2018_super,bryan_2019_excess}
which relies on Bayes' theorem to derive $P(\ILP|\OGP)$ from $P(\OGP|\ILP)$.

More recently, \citet{bonomo_2023_cold}
present a dedicated HARPS-N follow-up of 38 transiting \ILP hosts to search
for \OGPs. They report a lower but compatible rate of \OGP than in the general
population, in contradiction with \citet{zhu_2018_super} and
\citet{bryan_2019_excess}. They attribute this discrepancy to the possible
confusion between \OGPs and magnetic cycles or binary stars. Another possible
reason for this discrepancy, proposed by \citet{zhu_2024_metallicity}, is that
the stars observed by \citet{bonomo_2023_cold} are metal-poor compared to the
general population. However, the impact of stellar metallicity on the
(anti-)correlation between the two populations is still debated
\citep{bryan_2024_friends,vanzandt_2024_evidence,bryan_2025_resolving,vanzandt_2025_tesskeck,bonomo_2025_cold}.

More generally, the question of whether these populations correlate or anticorrelate is difficult to settle using archival data. Indeed, all of these studies
suffer from small-number statistics (small stellar sample sizes) and/or
nonhomogeneous samples, especially when comparing results from blind RV surveys
and transit follow-ups. As shown by \citet{vanzandt_2024_evidence}, some results
strongly depend on the exact definitions of \ILP and \OGP (in terms of
mass, radius, periods, and semi-major axis). This may partly reflect true
physical mechanisms, but could also arise from small-number statistics and biased
samples.

Additionally, a subtler issue, which we term ``observer-excitement''
bias, can affect some of these results. In many blind RV surveys, targets
showing promising signals tend to be monitored with higher
cadence, while others receive fewer observations. Consequently, 
blind surveys typically have more RV measurements (sometimes orders of
magnitude more) for targets with planets than for those without. This bias can have two relevant effects on \OGP-\ILP correlations
studies:
\begin{itemize}
  \item Low-amplitude planets are more likely to be detected in
        multi-planetary systems where an easier-to-detect planet has already
        been found, increasing the likelihood of intensive
        monitoring;
  \item When correcting for completeness, systems with planets typically have more points, and thus lower detection limits, giving them greater
        weight in occurrence estimates. This typically results in overestimation of occurrence rates.
\end{itemize}
This observer-excitement bias is very difficult to correct for, especially when combining blind surveys with different observing strategies.

Here, we present a long-term observational effort to study the \OGP-\ILP (anti-)correlation
by probing the \ILP occurrence in systems with and without \OGP.
We constructed two comparative samples of stars, with and without \OGP,
ensuring the homogeneity of the two samples
in terms of stellar properties and observing strategy, to avoid any bias that could affect the comparison.
In this first article of a series, we focus on the sample definitions
and on the first three detections (two new and one confirmation) of \ILP in \OGP hosts
around \targets{},
based on the first two years of observations of our program.
These three targets were known to host an \OGP
(see \citealp{tinney_2002_extrasolar} for \targetI{},
\citealp{marmier_2013_coralie} for \targetII{},
and \citealp{arriagada_2010_five} for \targetIII{}).
Moreover, a transiting \ILP has been detected by TESS (Transiting Exoplanet Survey Satellite) and confirmed with
follow-up PFS (Planet Finder Spectrograph) RV measurements \citep{teske2020} for \targetIII{}.

In Sect.~\ref{sec:sample}, we present the two samples and the observing strategy.
In Sect.~\ref{sec:stellar}, we present the properties of the three hosts,
and in Sect.~\ref{sec:observations} we describe the observations.
We present our analysis methods in Sect.~\ref{sec:analysis},
and our results for the three targets in Sect.~\ref{sec:results}.
Finally, in Sect.~\ref{sec:discussion}, we present preliminary conclusions from these detections. A statistical analysis of the full sample to derive the underlying occurrences of \ILP
with and without \OGP will be performed in subsequent articles.

\section{Sample definitions and observing strategy}
\label{sec:sample}

\begin{figure}
  \centering
  \includegraphics[scale=0.6]{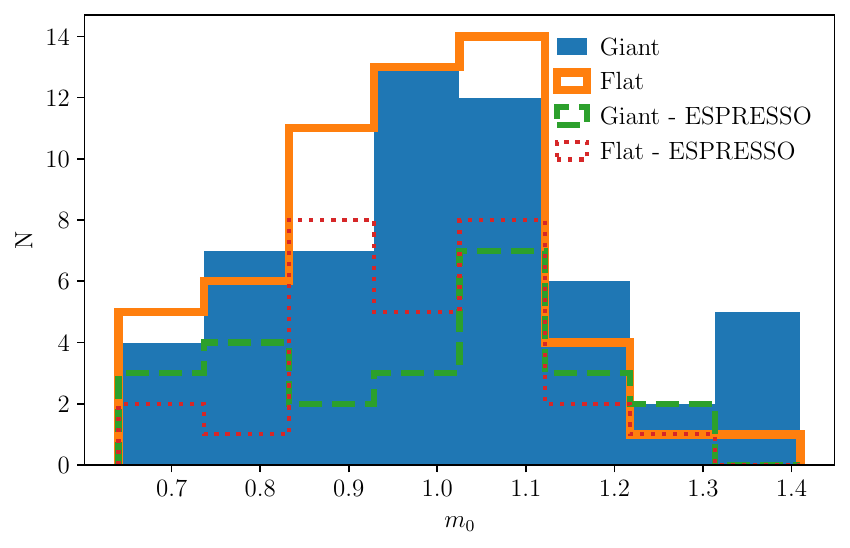}\\
  \hspace{2mm}\includegraphics[scale=0.6]{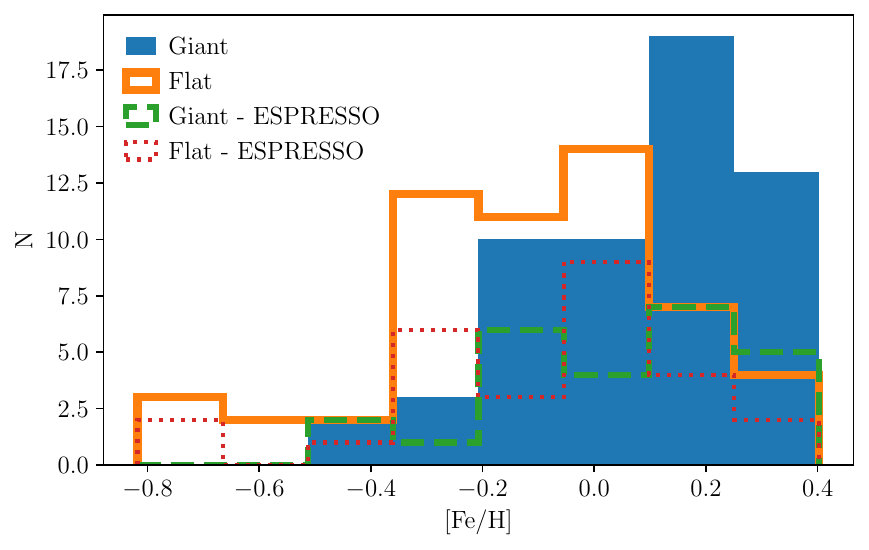}
  \caption{Masses (\textit{top panel}) and metallicities (\textit{bottom panel}) of the stars in our Giant and Flat samples (see Sect.~\ref{sec:sample}).
  }
  \label{fig:samples}
\end{figure}

We constructed our samples from two historical volume-limited giant-planet search surveys conducted with 
CORALIE \citep[]{udry_2000_coralie} and HARPS \citep{locurto_2010_harps} spectrographs.
These two historical surveys comprise a total of 2487 main-sequence
G and K dwarfs, volume-limited to 57~pc, and were conducted at a typical precision of about 5~m/s for CORALIE
and 2-3~m/s for HARPS over more than two decades.

We selected less active stars ($\log R'_{HK} < -4.6$) and considered only
stars closer than 52~pc. We defined two subsamples for our
comparative study, composed of stars with similar mass distributions ranging
from 0.65 to 1.4~$m_\odot$:
\begin{enumerate}
  \item ``Giant'' sample (all 57 systems matching the criteria):
        \begin{itemize}
          \item with at least one giant planet in their outer region ($m\sin i > 100~m_\oplus$, $a > 0.5$~AU),
          \item no giant planet in their inner region ($a < 0.5$~AU),
          \item no highly eccentric orbits ($e < 0.7$),
          \item no high-mass companions ($m < 30~m_J$ at 5~AU) based on Hipparcos-Gaia proper motion anomalies \citep[PMA, see][]{kervella_2022_stellar},
        \end{itemize}
  \item ``Flat'' sample (55 systems randomly selected among 73 systems matching the criteria):
        \begin{itemize}
          \item at least 25 HARPS and 25 CORALIE pre-existing measurements spread over at least 8 years,
          \item no signature of a giant planet from the analysis of available RVs,
          \item no signature of a giant planet from the analysis of Hipparcos-Gaia PMA.
        \end{itemize}
\end{enumerate}

We obtained two successive ESO Large Programmes (LPs) on HARPS (108.22KV) and
HARPS+ESPRESSO (112.25YG). In both campaigns, the aim was to explore the inner parts
of these systems by taking advantage of preexisting measurements.
We searched for \ILPs in the super-Earth (SE; 3--10~$m_\oplus$)
and Neptune-like (NE; 10--30~$m_\oplus$) ranges, with periods shorter than 50~d.

The goal of the first LP (108.22KV) was to obtain approximately 75 HARPS measurements for 26 systems belonging to the Giant sample.
The second LP (112.25YG), which is on-going at the time of writing, aims to gather 50 HARPS measurements for all 112 systems and
25 ESPRESSO measurements for half of the sample (25 systems from the Giant sample and 27 from the Flat sample).
Most systems already had HARPS measurements, with some having reached the 50, or even 75, measurement thresholds,
allowing reliance on archival data where appropriate.
We used exposure times of 15~minutes for HARPS and 12~minutes for ESPRESSO (split in sub-exposures for bright targets) to average out stellar oscillations \citep{dumusque_2011_planetary}.

We note that the two historical giant-planet search surveys were subject to the observer-excitement bias
described above.
Before our search for inner planets began,
the stars in the Giant sample had a median of 109 CORALIE measurements,
while the stars in the Flat sample had a median of 34 CORALIE measurements.
However, the detection capacity for low-mass inner planets in our samples is dominated
by the high-precision HARPS and ESPRESSO measurements,
for which we ensured a homogeneous observing strategy.
Moreover, these historical surveys are blind-search surveys (i.e., not transit follow-ups),
so there is no bias toward stars with known inner companions.
Thus, the observer-excitement bias in giant-planet search surveys could have increased the
probability of detecting additional \OGPs in systems that show a hint of \OGP presence.
However, the effect of this bias on the \OGP-\ILP correlation should be limited.

Fig.~\ref{fig:samples} shows the distribution of masses and metallicities of our two samples, as well as those of the sub-samples observed with ESPRESSO.
These masses and metallicities were computed using the ARES+MOOG method \citep[][see also Sect.~\ref{sec:stellar}]{sousa_sweet-cat_2021}.
The mass distributions of the two samples are very similar (Fig.~\ref{fig:samples}). 
Regarding metallicities, the Giant sample is biased toward higher metallicities
than the Flat sample, due to the metallicity-planet correlation \citep{gonzalez_1998_spectroscopic,santos_2001_metalrich}.
However, both samples contain a large number of targets in the range $-0.4 < [\mathrm{Fe/H}] < 0.2$, allowing an unbiased comparison (with respect to metallicity) by
excluding the most extreme targets in each sample.
In the longer term,
the Flat sample may be slightly refined for future observational programs
to exclude the most metal-poor stars
and include additional metal-rich stars that were
randomly excluded.

Here, we present three inner planet detections from our analysis of the 26
systems observed during the first LP (108.22KV). Global statistical analysis of
the sample will be performed in subsequent articles of this series.

\section{Stellar properties}
\label{sec:stellar}

\begin{table*}
  \renewcommand{\arraystretch}{1.3}
  \caption{Stellar parameters for \targetI, \targetII, and \targetIII.}
  \label{tab:stellar_parameters}
  \centering
  \begin{tabular}{lcccc}
    \hline \hline
    Star                      & \targetI                & \targetII              & \targetIII              &                      \\
    \hline
    Spectral Type             & F9.5V                   & G0V                    & G2V                     & \tablefootmark{a, b} \\

    $\pi$ (mas)               & 29.8633 $\pm$ 0.0194    & 25.0327 $\pm$ 0.0205   & 21.9301 $\pm$ 0.0267    & \tablefootmark{c}    \\
    $V$ (mag)                 & 7.11                    & 6.44                   & 7.93                    & \tablefootmark{a}    \\
    $B$ (mag)                 & 7.68                    & 7.05                   & 8.563                   & \tablefootmark{a}    \\
    \hline
    $d$ (pc)                  & 33.50$^{+0.05}_{-0.02}$ & 40.0$^{+0.05}_{-0.04}$ & 45.68$^{+0.09}_{-0.08}$ & \tablefootmark{d}    \\

    $\rm T_{eff} \, (K) $     & $5994^{+55}_{-64}$      & $6072^{+83}_{-76}$     & $6007^{+75}_{-71}$      & \tablefootmark{d}    \\
    $\rm [Fe/H] \, (dex) $    & $-0.129 \pm 0.041$      & $0.213 \pm 0.042$      & $-0.015 \pm 0.041$      & \tablefootmark{e}    \\
    $\log g \,$(cm\,s$^{-2}$) & 4.37 $\pm$ 0.02         & 4.09 $\pm 0.02$        & 4.41 $\pm 0.02$         & \tablefootmark{e, c} \\

    $R_* (R_{\odot})$         & 1.08$^{+0.02}_{-0.03}$  & 1.71$^{+0.03}_{-0.04}$ & 1.03$^{+0.02}_{-0.02}$  & \tablefootmark{d}    \\

    $M_* (M_{\odot})$         & 1.16$^{+0.02}_{-0.03}$  & 1.16$^{+0.13}_{-0.09}$ & 1.08$^{+0.05}_{-0.05}$  & \tablefootmark{d}    \\
    $L_{*} (L_{\odot})$       & 1.35$^{+0.08}_{-0.08}$  & 3.54$^{+0.27}_{-0.20}$ & 1.24$^{+0.09}_{-0.07}$  & \tablefootmark{d}    \\
    $\log R'_{\rm HK}$        & $-4.929 \pm 0.014$      & $-5.033 \pm 0.022$     & $-4.946 \pm 0.012$      & \tablefootmark{f}    \\
    $P_\mathrm{rot.}$ (d)     & $15.0 \pm 2.6$          & $22.5 \pm 2.9$         & $22.8 \pm 3.0$          & \tablefootmark{g}    \\
    Age (Gyr)                 & $2.6 \pm 1.2$           & $4.2 \pm 2.0$          & $3.8 \pm 1.8$           & \tablefootmark{g}    \\
  \end{tabular}
  \tablefoot{
    The uncertainties in this table only reflect the formal error propagation.
    For some parameters (e.g., the masses),
    the systematic errors in stellar models dominate these formal uncertainties
    \citep[e.g.,][]{tayar_2022_guide}.\\
    \tablefoottext{a}{\cite{gray_contributions_2006}};
    \tablefoottext{b}{\cite{houk_michigan_1988}};
    \tablefoottext{c}{\cite{riello_gaia_2021}};
    \tablefoottext{d}{Obtained from ARIADNE \citep{vines_ariadne_2022}};
    \tablefoottext{e}{Obtained using the ARES+MOOG method \citep{sousa_sweet-cat_2021}};
    \tablefoottext{f}{\cite{gomesdasilva_2021_stellar,rutten_1984_magnetic}};
    \tablefoottext{g}{\cite{mamajek_2008_improved}}.
  }
  \renewcommand{\arraystretch}{1}
\end{table*}

This section describes the fundamental parameters of \targets{}, which are part of
our Giant sample.
Obtaining accurate stellar masses is crucial,
as it directly impacts our estimates of the planetary (minimum) masses inferred from the RV time series.
Values of effective temperature $T_\text{eff}$, stellar mass $M_*$,
stellar luminosity $L_*$, stellar radius $R_*$, and distance $d$
were estimated using the ARIADNE (spectrAl eneRgy dIstribution bAyesian moDel averagiNg fittEr)
Python package \citep{vines_ariadne_2022}.

The ARIADNE package fits the spectral energy distribution (SED) of a star using multiple
stellar atmosphere model grids, including PHOENIX V2 \citep{husser_new_2013},
BT-Settl, BT-NextGen, BT-Cond \citep{allard_models_2012}, Castelli \& Kurucz
\citep{castelli_new_2003}, and Kurucz \citep{kurucz_atlas9_1993}. The package
automatically retrieves available broadband photometric data from various
sources and performs an SED fit using the different models. The resulting
fits are then compared using Bayesian model averaging, which yields estimates of the stellar parameters. All of the results presented here were obtained by
running ARIADNE with default priors \citep[see Table~2 of][]{vines_ariadne_2022}.

Because the SED does not precisely constrain the stellar metallicity [Fe/H] and
surface gravity $\log g$, we derived these values
using the ARES+MOOG method \citep{sneden_carbon_1973,sousa_new_2007,sousa_2008_spectroscopic,sousa_ares_2014,sousa_2015_ares,sousa_sweet-cat_2021}.
Additionally, surface
gravity values were obtained from the Gaia Early Data Release 3
\citep{gaia_collaboration_gaia_2016,riello_gaia_2021} parallaxes and photometry.
We computed the $\log R'_{\rm HK}$ time series using \cite{gomesdasilva_2021_stellar,rutten_1984_magnetic}
and derived estimates of the rotation period and age using \citet{mamajek_2008_improved}.
A summary of the derived stellar parameters is given in
Table~\ref{tab:stellar_parameters}.

\paragraph{\targetI}
is an F9.5\,V dwarf at a distance $d$ = 33.5~pc with a magnitude of
$V$ = 7.11. The ARIADNE fit yields an effective temperature $T_\text{eff}$ =
5994~K and a mass estimate $M_*$ = 1.16~M$_\odot$. The star has a surface gravity $\log g$ = 4.37 cm\,s$^{-2}$ and a metallicity [Fe/H] = -0.129~dex.

\paragraph{\targetII}
is a G0\,V star with magnitude $V$ = 6.44. From the ARIADNE
analysis, we find that this star is located at a distance $d$ = 40~pc, has
an effective temperature $T_\text{eff}$ = 6072~K, and a mass $M_*$ = 1.16~M$_\odot$.
The surface gravity and metallicity are $\log g$ = 4.09 cm\,s$^{-2}$ and [Fe/H] = 0.213~dex,
respectively.

\paragraph{\targetIII}
is a G2\,V dwarf with magnitude $V$ = 7.93 located at a distance $d$ = 45.68~pc. From the SED fit, we obtain an effective temperature
$T_\text{eff}$ = 6007 K. This star has an estimated mass $M_*$ = 1.08~M$_\odot$,
a metallicity [Fe/H] = -0.015~dex, and a surface gravity $\log g$ =
4.41 cm\, s$^{-2}$.

\section{Observations}
\label{sec:observations}

This section describes the combined RV time series for the three targets from
CORALIE and HARPS observations, with the addition of PFS observations for \targetIII{}.

The CORALIE instrument is an echelle spectrograph installed on the Euler Swiss telescope at La
Silla Observatory in Chile since 1998. It underwent upgrades in 2007 and
2014, which led to possible RV offsets. We consider the three versions of
CORALIE as different instruments: CORALIE98, CORALIE07, and CORALIE14.
Accordingly, we adjusted for a different offset and a different jitter term for each CORALIE version.
The HARPS instrument is
also an echelle spectrograph at La Silla Observatory, installed on the 3.6m
telescope since 2003. We distinguish two versions of HARPS as different
instruments, HARPS03 and HARPS15, corresponding to before and after a fiber link upgrade
performed in May 2015 \citep{locurto_harps_2015}.

For \targetIII{}, we also used the PFS RV measurements from \citet{teske2020}.
The PFS instrument is an echelle spectrograph
installed on the 6.5m Magellan II telescope at Las Campanas Observatory in
Chile. Its CCD was upgraded in 2018, and we distinguish two versions of the
instrument, PFS1 and PFS2, corresponding to before and after this upgrade. We used these PFS RVs
in addition to the CORALIE and HARPS time series.

A summary of the observations used in the analysis of the three targets is
given in Table~\ref{tab:observations}. The table lists both the number of spectra
and the number of unique nights of observations for each instrument. In the
following, we used nightly binned time series.

\begin{table}
  \caption{Summary of observations used in our analysis.}
  \label{tab:observations}
  \centering
  \setlength{\tabcolsep}{2pt}
  \begin{tabular}{llcccccc}
    \hline \hline
            &         & \multicolumn{2}{c}{HD 23079} & \multicolumn{2}{c}{HD 196067} & \multicolumn{2}{c}{HD 86226}                               \\
            &         & \#                           & avg. err.                     & \#                           & avg. err. & \#  & avg. err. \\
    \hline
    CORALIE & spectra & 129                          & 3.89                          & 116                          & 4.02      & 116 & 4.41      \\
            & nights  & 123                          & 3.88                          & 112                          & 3.83      & 114 & 4.34      \\
    HARPS   & spectra & 77                           & 0.67                          & 76                           & 0.66      & 74  & 0.84      \\
            & nights  & 76                           & 0.67                          & 75                           & 0.63      & 73  & 0.83      \\
    PFS     & spectra & --                           & --                            & --                           & --        & 105 & 1.4       \\
            & nights  & --                           & --                            & --                           & --        & 59  & 1.34
  \end{tabular}
\end{table}

\section{Analysis}
\label{sec:analysis}

We performed two analyses for the three systems: a simple periodogram
analysis (Sect.~\ref{sec:kepmodel}) and posterior inference using a nested
sampling approach (Sect.~\ref{sec:kima}).

\subsection{Periodogram analysis}
\label{sec:kepmodel}

As an initial, simple, and fast analysis, we performed a period search using the
open-source \texttt{kepmodel}\footnote{Available at:
  \url{https://gitlab.unige.ch/delisle/kepmodel}} Python package
\citep{delisle_2016_analytical,delisle_2020_efficient}. We first defined a base
model of the RV time series, with an offset and a jitter term for each
instrument. The jitter term is added quadratically to the RV uncertainties. We
performed a maximum-likelihood fit of the model, restricting the jitter terms to
be lower than 20 \ms (see Table~\ref{tab:priors}). We then searched sequentially
for planetary signals in the residuals of the RV time series by computing a
periodogram and the associated false alarm probability \citep[FAP;
  see][]{baluev_2008_assessing,delisle_2020_efficient}. For low FAP values (below 0.1\%),
we selected the highest peak in the periodogram and computed an estimate of the
planet's orbital parameters
\citep[see][]{delisle_2016_analytical,delisle_2022_analytical}. We then
performed a global maximum likelihood fit, adjusting all planetary parameters as
well as the instruments' offsets and jitter terms. We iterated the procedure
until no significant periodic signal was found in the residuals (i.e.,
$\mathrm{FAP} > 0.1\%$).

\subsection{Posterior sampling}
\label{sec:kima}

To better determine the orbital architecture of the systems and estimate
uncertainties on the orbital parameters, we explored the parameter space using a
nested sampling approach. The same model was used for the three targets, but
with different prior distributions for the outer companions. Since the
detection of these outer companions is not in question, we set some
priors based on the previous periodogram analysis. This choice does not 
significantly impact the resulting posterior distributions. For the inner part
of the systems, we adopted weakly informative priors, since the goal was to detect
lower-mass planets, if present.

The model then consists of the following components: a Keplerian function for
the outer companion, up to two additional Keplerians for the inner part of the
system, between-instrument RV offsets, a constant systemic velocity, and a
Gaussian noise model per instrument that includes individual jitter terms. The
prior distributions for the parameters are listed in Table~\ref{tab:priors}.

We used the Gaia DR3 radial velocity measurement \citep{katz2023} to set the
prior on the systemic velocity, $v_{\rm sys}$, and the associated uncertainty to
assign a Gaussian prior for the between-instrument RV offsets. Even if
relatively informative, these priors are much wider than the posteriors and do
not influence the results. For the jitter terms, we assigned a modified
log-uniform prior \citep[see e.g.][]{gregory2005} with a knee at 1 \ms and an
upper limit of 20 \ms. The number of additional Keplerians, $N_p$, was free in
this analysis and was assigned a uniform prior from 0 to 2, motivated by the fact
that the periodogram analysis from section \ref{sec:kepmodel} showed only one
inner companion in each system. The orbital parameters of these two Keplerians
share the same priors: a log-uniform distribution for the periods, between 1 day
and the time span of the time series, a modified log-uniform distribution for
the semi-amplitudes (with a knee at 1 \ms and an upper limit of 100 \ms), a
Kumaraswamy distribution \citep{kumaraswamy1980} for the eccentricities,
and uniform priors for the argument of periastron, $\omega$,
and the mean anomaly, $M_0$.
The Kumaraswamy distribution is very similar to the beta distribution proposed by \citet{kipping2013},
but is easier to implement in nested sampling algorithms since its inverse cumulative distribution function
has a closed form \citep[see also][]{Stevenson2025}.

\begin{table}
  \renewcommand{\arraystretch}{1.4}
  \caption{Prior distributions used in the RV analysis.}
  \label{tab:priors}
  \centering
  \begin{tabular}{l c l l}
    \hline\hline
    Parameter           & Units & Prior                                                                                  \\
    \hline
    \multicolumn{3}{l}{general}                                                                                          \\
    $v_{\rm sys}$       & \ms   & $\mathcal{G}$\,(RV$_{\rm Gaia}$, $\sigma$RV$_{\rm Gaia}$)                              \\
    $v_{\rm offset, i}$ & \ms   & $\mathcal{G}$\,(0, $\sigma$RV$_{\rm Gaia}$)                                            \\
    $j$                 & \ms   & $\mathcal{MLU}$\,(1, 20)                                                               \\
    \multicolumn{3}{l}{outer companion}                                                                                  \\
    $P$                 & days  & $\mathcal{LU} \left( 0.8 \!\cdot\! P_{\rm guess}, 1.2 \!\cdot\! P_{\rm guess} \right)$ \\
    $K$                 & \ms   & $\mathcal{U} \left( 0.5 \!\cdot\! K_{\rm guess}, 1.5 \!\cdot\! K_{\rm guess} \right)$  \\
    $e$                 &       & $\mathcal{U} \left( 0, 1 \right)$                                                      \\
    $\omega$            & rad   & $\mathcal{U} \left( 0, 2\pi \right)$                                                   \\
    $M_0$               & rad   & $\mathcal{U} \left( 0, 2\pi \right)$                                                   \\
    \multicolumn{3}{l}{additional Keplerians}                                                                            \\
    $N_p$               & days  & $\mathcal{U}_i \left( 0, 2 \right)$                                                    \\
    $P$                 & days  & $\mathcal{LU}  \left( 1, \Delta t \right)$                                             \\
    $K$                 & \ms   & $\mathcal{MLU} \left( 1, 100 \right)$                                                  \\
    $e$                 &       & $\mathcal{K}   \left( 0.867, 3.03 \right)$                                             \\
    $\omega$            & rad   & $\mathcal{U}   \left( 0, 2\pi \right)$                                                 \\
    $M_0$               & rad   & $\mathcal{U}   \left( 0, 2\pi \right)$                                                 \\
    \hline
  \end{tabular}
  \tablefoot{The distributions are abbreviated as follows:
    $\mathcal{G}$~=~Gaussian with mean and standard deviation;
    $\mathcal{MLU}$~=~modified log-uniform with knee and upper limit;
    $\mathcal{LU}$~=~log-uniform with lower and upper limits;
    $\mathcal{U}$~=~uniform with lower and upper limits;
    $\mathcal{K}$~=~Kumaraswamy with parameters $a$ and $b$.
    $P_{\rm guess}$ and $K_{\rm guess}$ denote the maximum likelihood
    estimates for the period and semi-amplitude of the outer
    companion.
    $\Delta t$ is the time span of the time series.}
  \renewcommand{\arraystretch}{1}
\end{table}

We sampled from the posterior distribution using \texttt{kima} \citep{kima}, a
package\footnote{Available at: \url{https://github.com/kima-org/kima}} dedicated
to the Bayesian analysis of RV time series. The code uses the diffusive nested
sampling algorithm \citep{brewer2011} to explore the posterior and calculate the
marginal likelihood, or evidence, of the model. The algorithm is a variant of
classic nested sampling \citep{skilling2006} with some unique advantages. It explores a mixture of nested distributions, each successively occupying
smaller regions within the prior, guided by the likelihood function. It can
handle distributions showing multimodality, strong parameter correlations, and
phase transitions \citep[see e.g.][]{buchner2023}, much like those typically
encountered in the exoplanet detection problem
\citep[e.g.][]{ford2005a,gregory2005,feroz2011a}. Moreover, the algorithm can
handle so-called trans-dimensional models \citep{brewer2014,brewer2015}, in which
the number of parameters can change depending on an unknown number of
components (Keplerian functions, in our case).
We ran the algorithm for a fixed number of steps ($500\,000$), which guaranteed
a sufficient effective sample size (ESS).

\section{Results}
\label{sec:results}

\begin{figure*}
  \centering
  {\footnotesize
    \setlength{\tabcolsep}{0pt}
    \begin{tabular}{lll}
      \hspace{1cm}\targetI{}                                 &
      \hspace{1cm}\targetII{}                                &
      \hspace{1cm}\targetIII{}                                 \\
      \includegraphics[width=0.33\linewidth]{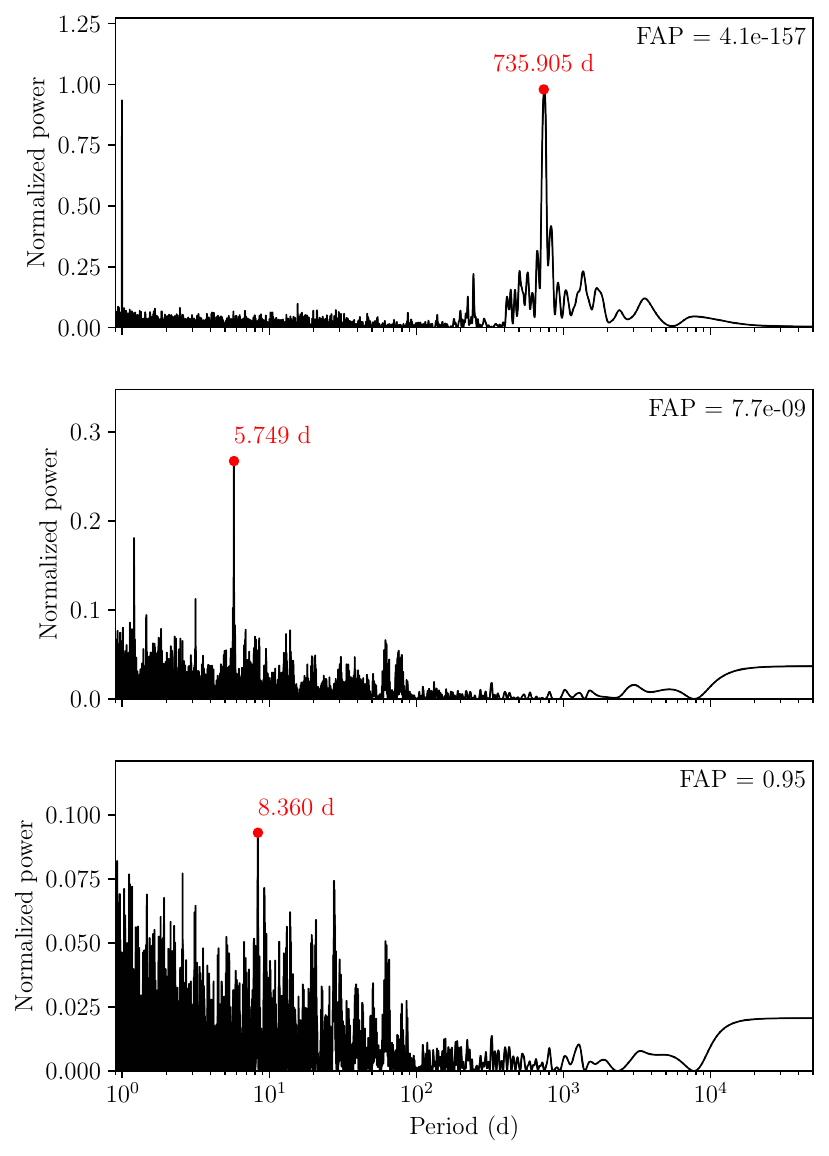}  &
      \includegraphics[width=0.33\linewidth]{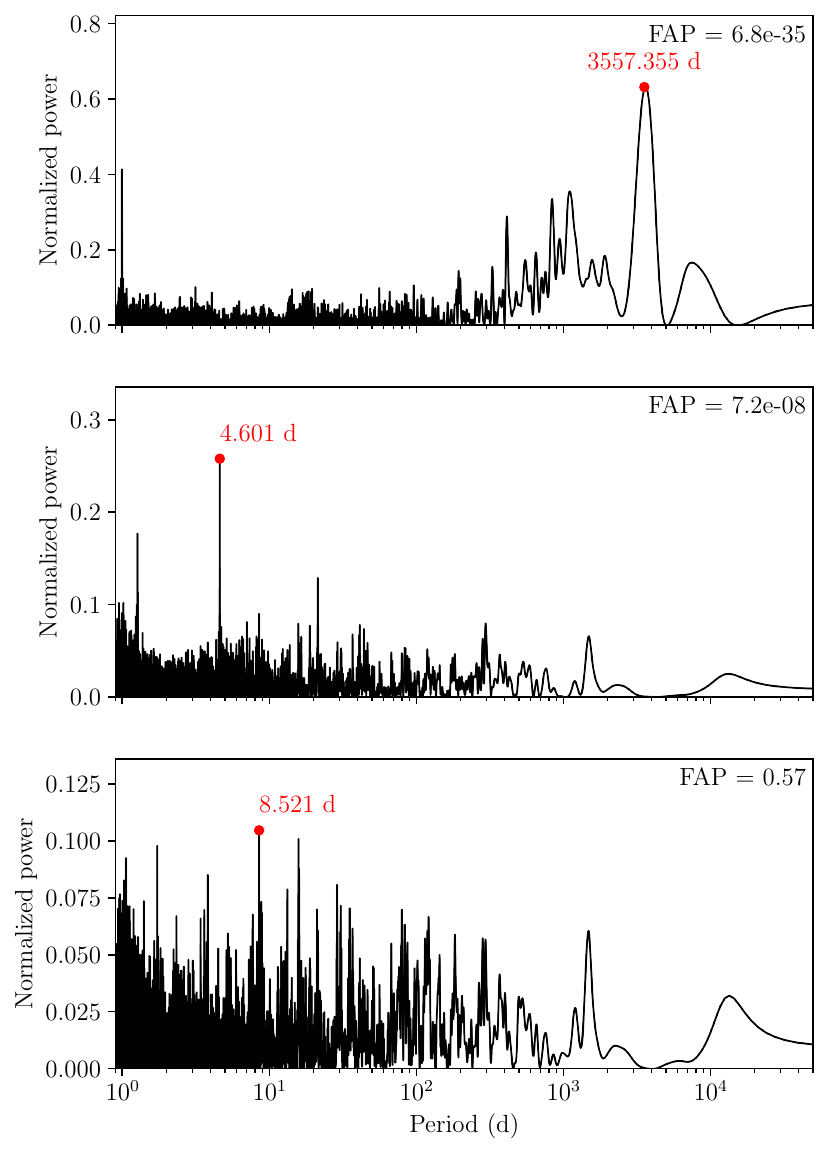} &
      \includegraphics[width=0.33\linewidth]{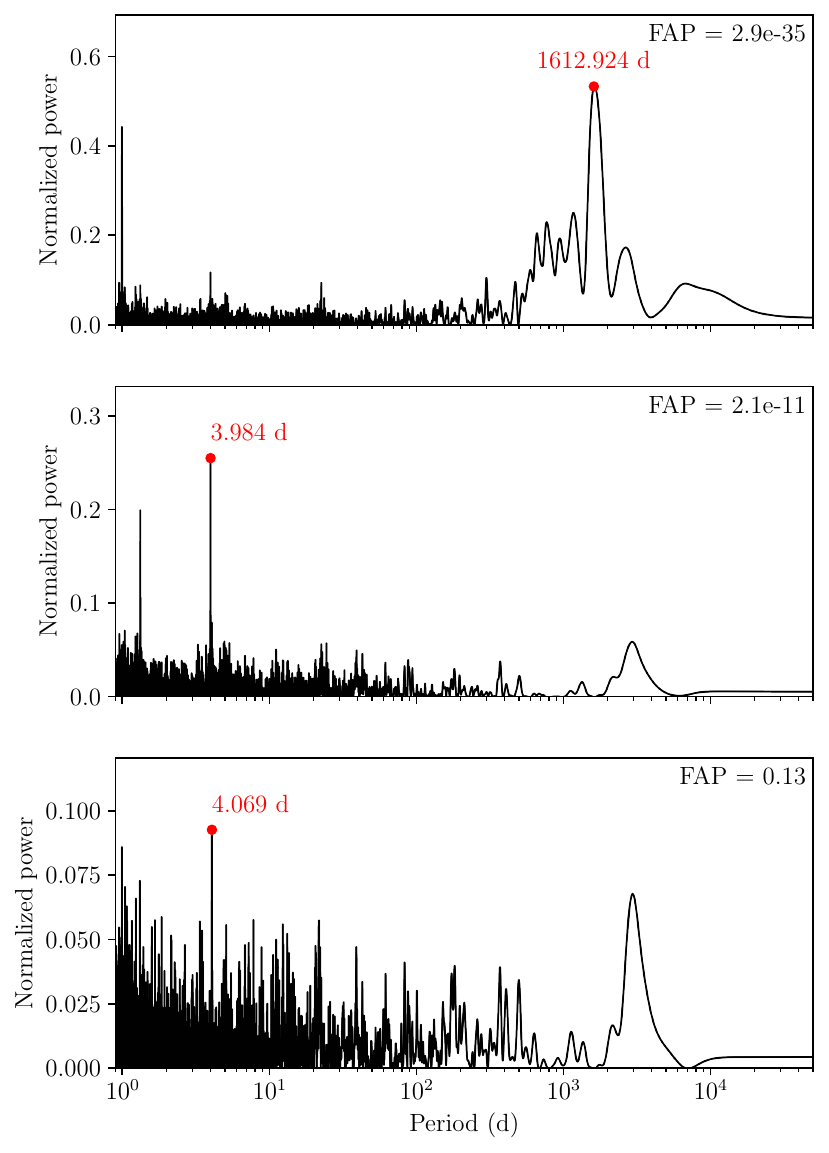}
    \end{tabular}
  }
  \caption{Periodograms of the raw RV time series (\textit{top panel})
    as well as of the residuals after subtracting the \OGP (\textit{middle panel})
    and the \ILP (\textit{bottom panel}), for
    \targetI{} (\textit{left}),
    \targetII{} (\textit{center}), and
    \targetIII{} (\textit{right}).
  }
  \label{fig:periodograms}
\end{figure*}

\begin{figure*}
  \centering
  \includegraphics[width=0.33\hsize]{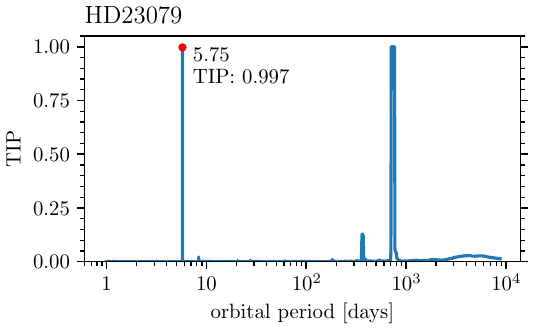}
  \includegraphics[width=0.33\hsize]{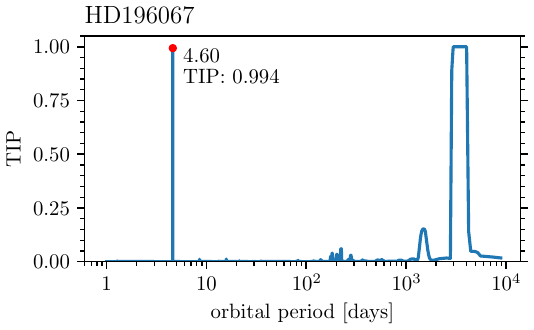}
  \includegraphics[width=0.33\hsize]{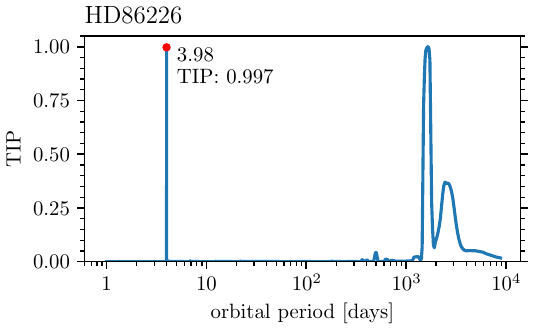}
  \caption{Posterior distributions for the orbital periods from the analysis
    of the three targets (left: \targetI, center: \targetII, right:
    \targetIII). The plots show the true inclusion probability (TIP),
    highlighting the peaks corresponding to the detected inner planets.
    Because the Keplerian signal of the outer planet is always included
    in the model, the TIP value for the outer planet period is, by
    definition, unity.}
  \label{fig:tip}
\end{figure*}

\begin{table}
  \centering
  \renewcommand{\arraystretch}{1.5}
  \setlength{\tabcolsep}{1pt}
  \caption{Posteriors of the orbital parameters for \targetI, \targetII, and \targetIII.}
  \label{tab:final_params_short}
  {\footnotesize
    \begin{tabular}{l c c | c | c}
      \hline
      \hline
      Parameter                                                                       & unit                              & \targetI                          & \targetII   & \targetIII  \\
      \hline
      \hline
      $P_\mathrm{b}$                                                                  & d                                 & $735.7 ^{+0.4} _{-0.4}$
                                                                                      & $3397 ^{+17} _{-14}$
                                                                                      & $1616 ^{+46} _{-28}$
      \\
      $U_\mathrm{0,\,b}${\tiny$=M_\mathrm{0,\,b}+\omega_\mathrm{b}$}\tablefootmark{a} & deg                               & $107.0 ^{+0.9}_{-0.9}$
                                                                                      & $-53.0 ^{+2.6} _{-2.2}$
                                                                                      & $-66 ^{+16} _{-12}$
      \\
      $K_\mathrm{b}$                                                                  & \ms                               & $54.4 ^{+0.3} _{-0.4}$
                                                                                      & $80 ^{+3} _{-4}$
                                                                                      & $7.8 ^{+0.8} _{-0.6}$
      \\
      $e_\mathrm{b}$                                                                  &                                   & $0.072 ^{+0.012} _{-0.011}$
                                                                                      & $0.561 ^{+0.011} _{-0.012}$
                                                                                      & $0.07 ^{+0.09} _{-0.05}$
      \\
      $\omega_\mathrm{b}$\tablefootmark{b}                                            & deg                               & $29 ^{+15} _{-7}$
                                                                                      & $142.0 ^{+1.5} _{-1.8}$
                                                                                      & $37 ^{+100} _{-180}$
      \\
      $m_b\sin i_b$\tablefootmark{c}                                                  & $M_J$                             & $2.66 ^{+0.05}_{-0.05}$
                                                                                      & $5.42 ^{+0.20} _{-0.26}$
                                                                                      & $0.49 ^{+0.05} _{-0.04}$
      \\
      \hline
      $P_\mathrm{c}$                                                                  & d                                 & $5.74894 ^{+0.00022} _{-0.00021}$
                                                                                      & $4.60111^{+0.00024}_{-0.00024}$
                                                                                      & $3.98450 ^{+0.00023} _{-0.00022}$
      \\
      $U_\mathrm{0,\,c}${\tiny$=M_\mathrm{0,\,c}+\omega_\mathrm{c}$}\tablefootmark{a} & deg                               & $156 ^{+10}_{-10}$
                                                                                      & $-1 ^{+18} _{-17}$
                                                                                      & $-10 ^{+24} _{-22}$
      \\
      $K_\mathrm{c}$                                                                  & \ms                               & $2.65 ^{+0.26} _{-0.27}$
                                                                                      & $3.6 ^{+0.4}_{-0.4}$
                                                                                      & $2.82 ^{+0.25} _{-0.26}$
      \\
      $e_\mathrm{c}$                                                                  &                                   & $0.07 ^{+0.09} _{-0.05}$
                                                                                      & $0.12 ^{+0.12} _{-0.09}$
                                                                                      & $0.05 ^{+0.07} _{-0.04}$
      \\
      $\omega_\mathrm{c}$\tablefootmark{b}                                            & deg                               & $247 ^{+64} _{-166}$
                                                                                      & $278 ^{+41} _{-104}$
                                                                                      & $170 ^{+108} _{-94}$
      \\
      $m_c\sin i_c$\tablefootmark{c}                                                  & $M_\oplus$                        & $8.1 ^{+0.8}_{-0.8}$
                                                                                      & $10.1 ^{+1.2}_{-1.2}$
                                                                                      & $7.7 ^{+0.7}_{-0.7}$                                                                              \\
      \hline
      \hline
      Epoch                                                                           & BJD                               & 2\,455\,000                       & 2\,455\,000 & 2\,455\,000
    \end{tabular}
  }
  \tablefoot{
    An extended version of this table with the maximum likelihood estimates and \texttt{kima} posteriors of all model parameters is provided in Table~\ref{tab:final_params}.\\
    \tablefoottext{a}{Mean argument of latitude at the reference epoch ($U_0=M_0 + \omega=\lambda_0-\Omega$, where $M_0$, $\lambda_0$, $\omega$, and $\Omega$ are the mean anomaly, mean longitude, argument of periastron, and longitude of the ascending node, respectively).
    }
    \tablefoottext{b}{Argument of periastron of the stellar orbit.
    }
    \tablefoottext{c}{The uncertainties on the planetary masses take into account the stellar mass formal uncertainties, but do not include systemic errors from stellar models (see Table~\ref{tab:stellar_parameters}).
    }}
  \renewcommand{\arraystretch}{1}
\end{table}

\begin{figure}
  {\footnotesize
    \indent\hspace{1cm}\targetI{}\\
    \includegraphics[width=0.95\linewidth]{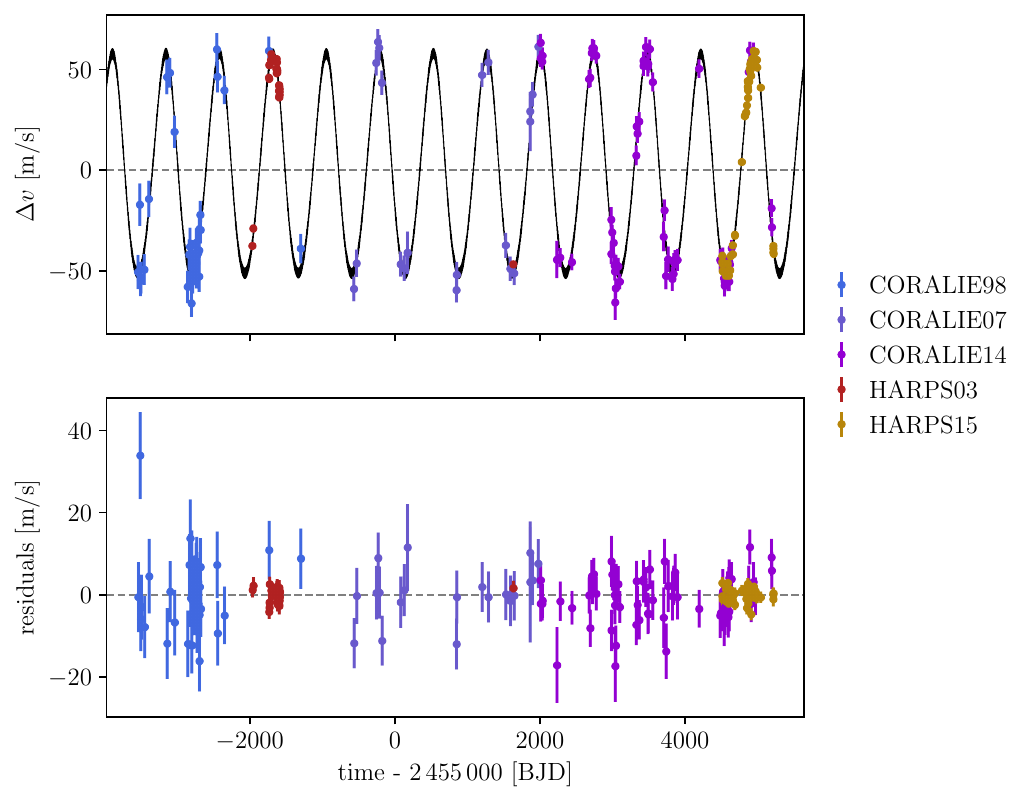}\\
    \indent\hspace{1cm}\targetII{}\\
    \includegraphics[width=0.95\linewidth]{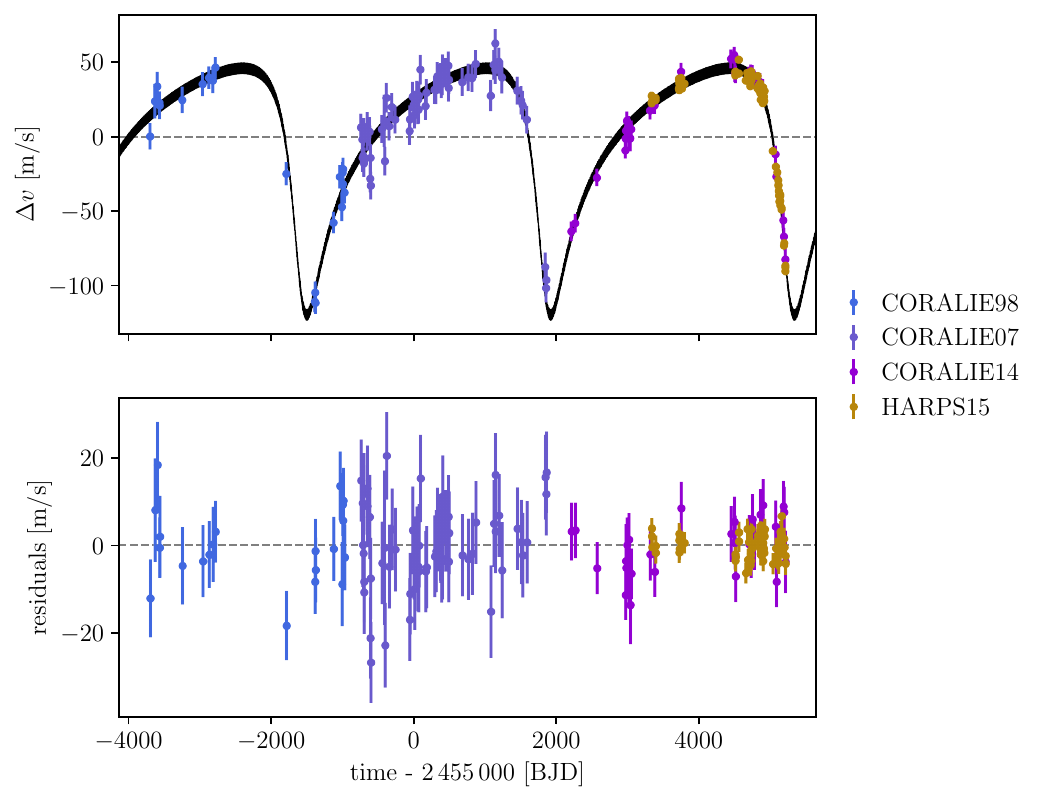}\\
    \indent\hspace{1cm}\targetIII{}\\
    \includegraphics[width=0.95\linewidth]{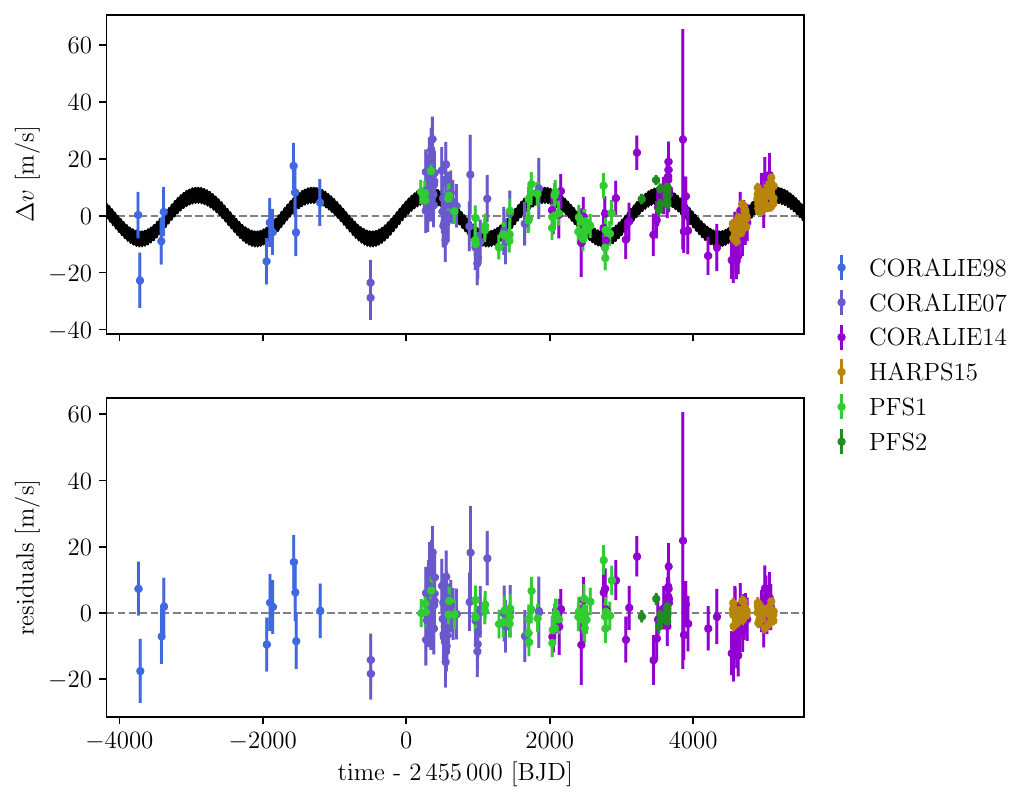}
  }
  \caption{Time series (\textit{top panel}) and residuals (\textit{bottom panel}) of the maximum-likelihood fit for each target (\targets{}).
  }
  \label{fig:timeseries}
\end{figure}

\begin{figure*}
  \centering
  {\footnotesize
    \begin{tabular}{l}
      \hspace{1cm}\targetI{}                                        \\
      \includegraphics[width=0.75\linewidth]{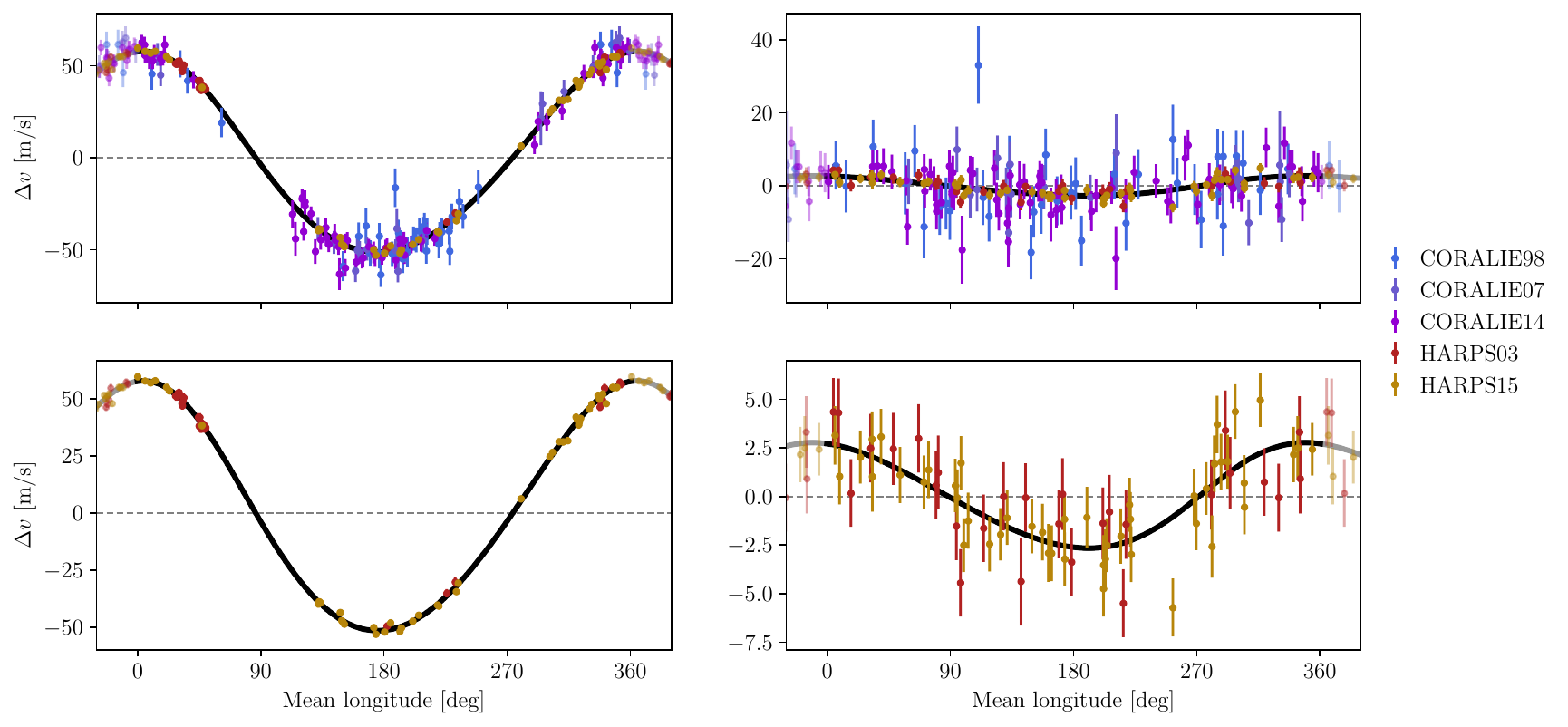}  \\
      \hspace{1cm}\targetII{}                                       \\
      \includegraphics[width=0.75\linewidth]{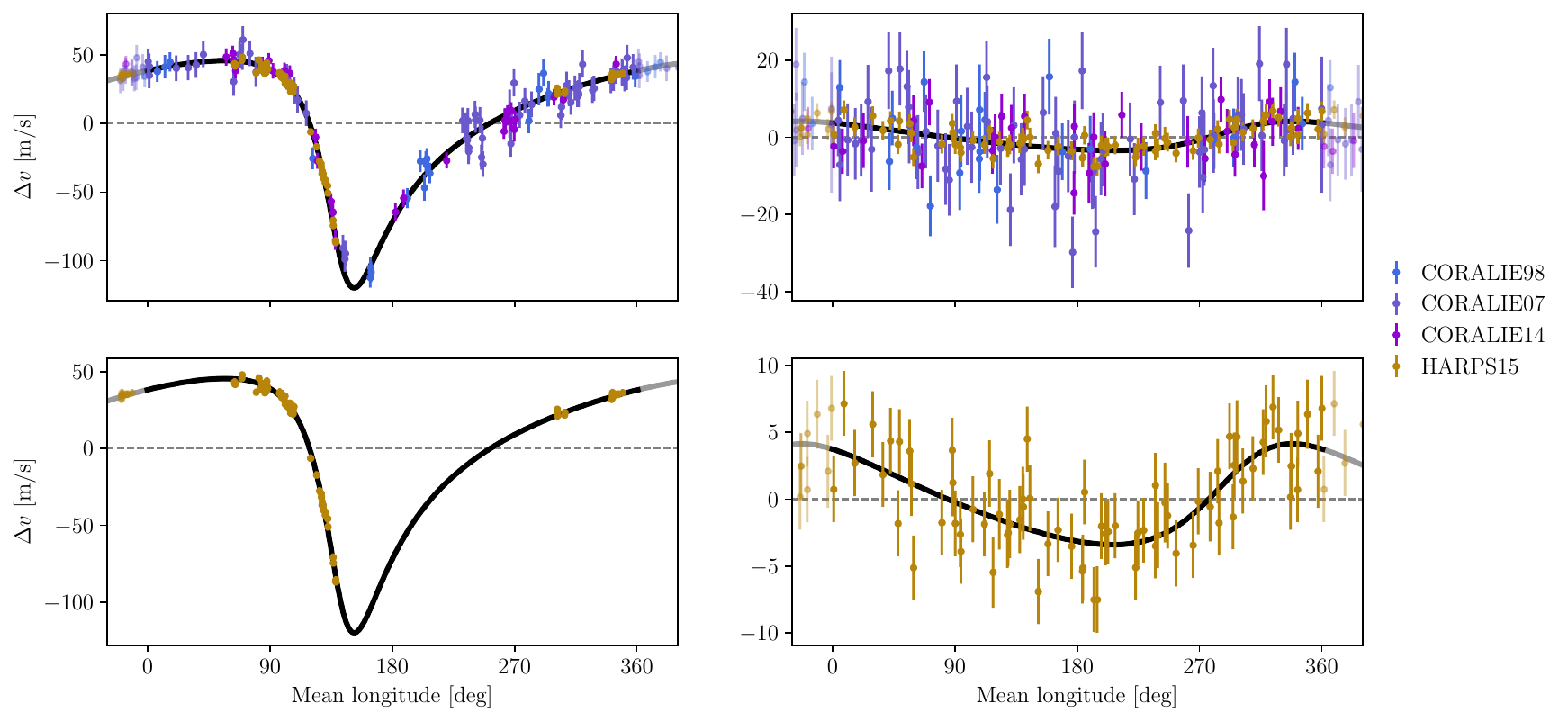} \\
      \hspace{1cm}\targetIII{}                                      \\
      \includegraphics[width=0.75\linewidth]{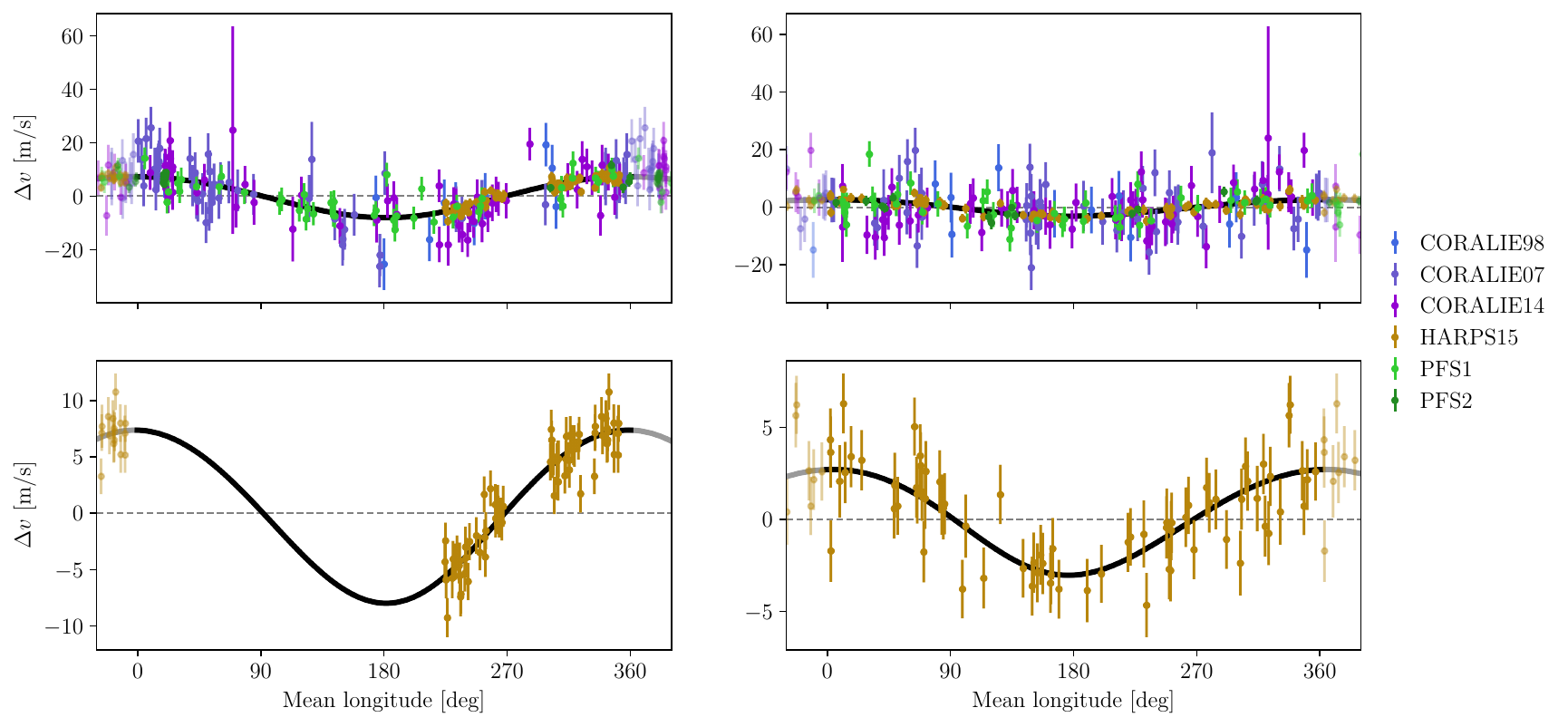}
    \end{tabular}
  }
  \caption{
    \textit{Top panel}: Residuals of the maximum-likelihood fit phase-folded
    at the periods of the two detected planets, showing the outer giant
    (\textit{left}) and lower-mass inner planet (\textit{right}) for each target (\targets{}).
    \textit{Bottom panel}: Same residuals, but showing only the HARPS data.
  }
  \label{fig:phasefolds}
\end{figure*}

\begin{figure*}
  \centering
  \includegraphics[width=0.33\hsize]{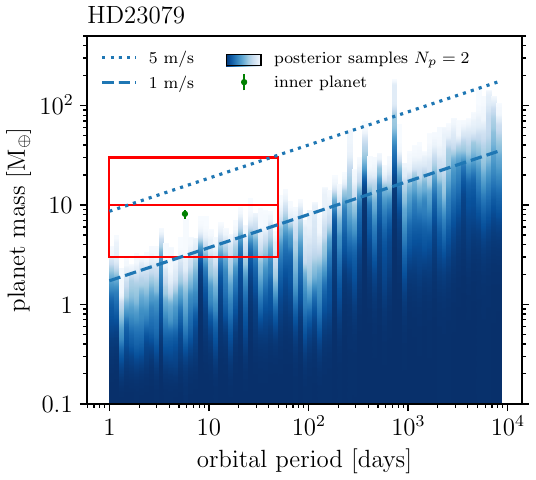}
  \includegraphics[width=0.33\hsize]{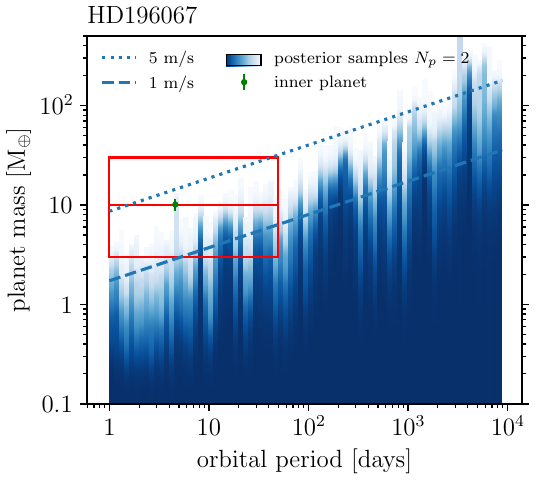}
  \includegraphics[width=0.33\hsize]{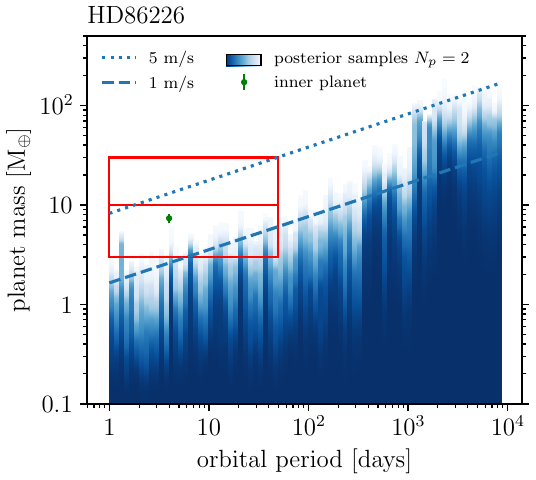}
  \caption{Compatibility limits from the analysis of the three targets (\textit{left}:
    \targetI, \textit{center}: \targetII, \textit{right}: \targetIII). The plots show a
    color map of the posterior density for planet mass in orbital period
    bins, for all samples with $N_p=2$ (i.e., with a total of three Keplerians, including the \OGP). Green points indicate
    detected inner planets with corresponding mass uncertainties.
    Dashed and dotted blue lines indicate masses corresponding to 1
    and 5 m/s semi-amplitude Keplerians, respectively. Masses are
    calculated using the stellar masses from Table \ref
    {tab:stellar_parameters}.
    Red boxes highlight the SE (3--10~$m_\oplus$)
    and NE (10--30~$m_\oplus$) mass ranges within 50~d.}
  \label{fig:detlim}
\end{figure*}

\begin{figure}
  \centering
  \includegraphics[width=\linewidth]{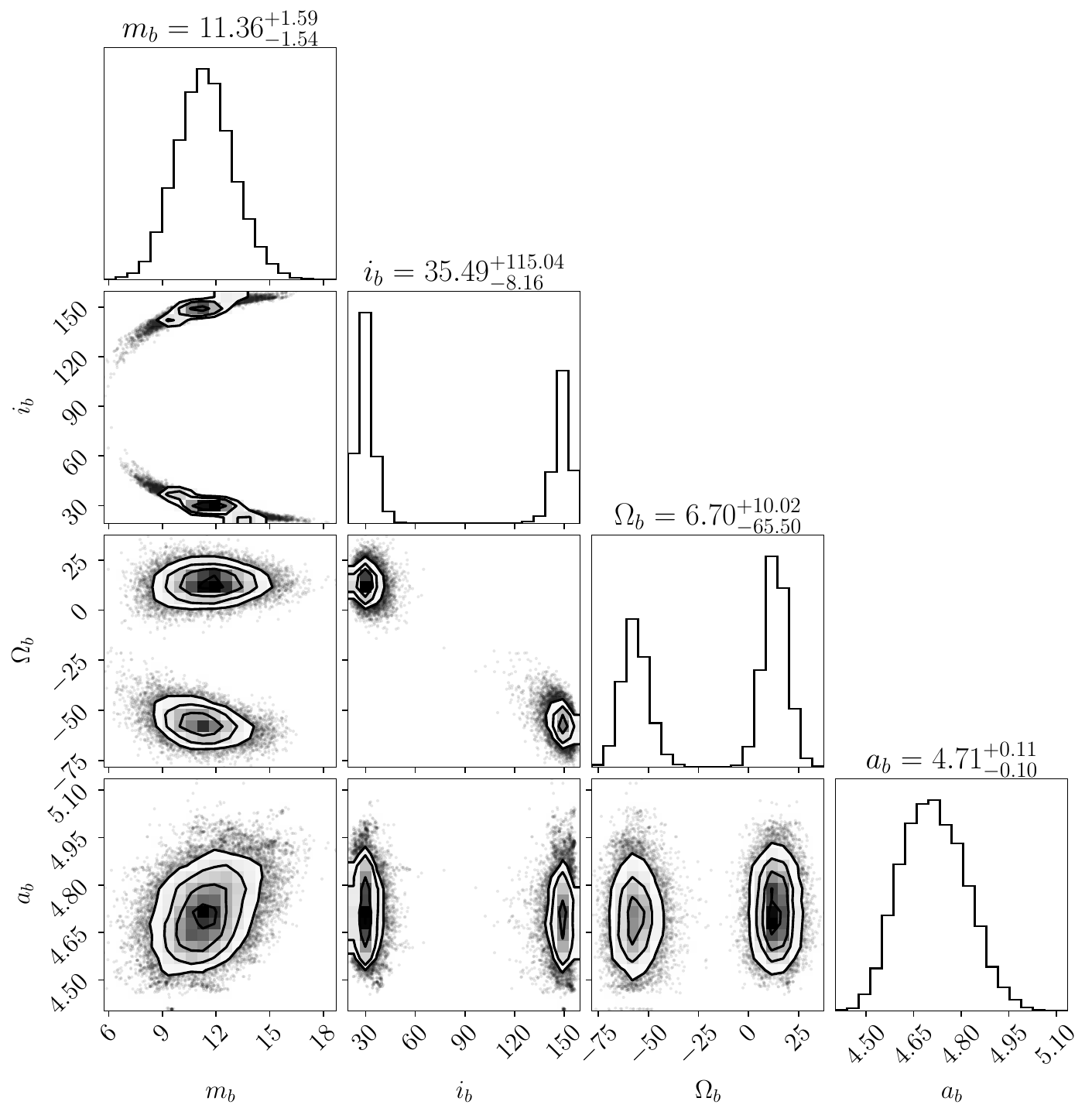}
  \caption{Posterior distributions from the \texttt{orvara} fit of \targetII{} (see Sect.~\ref{sec:results}).}
  \label{fig:HD196067_orvara}
\end{figure}

Fig.~\ref{fig:periodograms} shows successive periodograms of the raw RV
time series and after subtracting planetary signal for the three targets
(\targets). In all three systems, the periodogram analysis clearly indicates the
detection of two planets, an \OGP and an \ILP. These detections are significant,
as the FAP of all six planet signals are lower than $10^{-7}$. The FAP of the
residuals after subtracting the planetary signals are above 10\% in all three
time series, indicating that no clear signal remains unmodeled in the residuals.

We show the true inclusion probability (TIP) periodograms based on the
\texttt{kima} posterior sampling in Fig.~\ref{fig:tip}.
The TIP
\citep[see][]{hara_2022_detecting} is the posterior
probability (conditioned on the data) of having at least one planet in a small
period interval around a given period $P$, or equivalently, in a small frequency
interval $\left[f - \Delta f/2, f+\Delta f/2\right]$ around a frequency $f=1/P$.
 The TIP has been mathematically shown to be an optimal exoplanet detection criterion,
in the sense that it maximizes the expected number of true detections
for a given tolerance to false detections \citep[see][]{hara_2024_continuous}.
The TIP
periodograms of Fig.~\ref{fig:tip} were obtained using $\Delta f =
  1/T_\mathrm{span}$ (where $T_\mathrm{span}$ is the time span of the
observations) and by scanning the frequency $f$ over a finer grid ($\Delta
  f/10$) to ensure that local maxima of the TIP were not missed.
The TIP is marginalized over the number of planets included in the model.
In the \texttt{kima}
runs, we always included the \OGP in the model and allowed the number of additional planets to vary between zero and two
(thus the total number of planets is free between one and three).
Therefore, by construction, the TIP is one at the \OGP's period
because we enforced the presence of this already known planet in the model.
For the \ILP, the TIP values exceed 99\% for all
three targets (see Fig.~\ref{fig:tip}), which confirms the detection of these
planets. As in the classical periodograms, we do not find any additional
significant peak in the TIP periodograms.

The parameters of the maximum-likelihood solution and of the posterior
sampling are given in Table~\ref{tab:final_params} for all three systems.
Table~\ref{tab:final_params_short} provides a shorter version of this table that contains only the posteriors of the orbital parameters.
The two estimates are in good agreement for all parameters.
As final estimates, we quote the maximum-likelihood values and
the 68\% credible intervals from the posteriors. The RV time series and the
residuals from the maximum-likelihood solution are shown in
Fig.~\ref{fig:timeseries} and are phase-folded in Fig.~\ref{fig:phasefolds}.
Our solution for \targetIII{} agrees with \citet{teske2020},
with a slight refinement to the mass estimate for the inner planet~(c).

From the \texttt{kima} analysis, we also derived so-called compatibility limits,
corresponding to the orbital periods and masses of undetected planets that remain compatible with the available data (see, e.g., \citealt{standing2022} and
\citealt{figueira2025}). In practice, these limits were obtained from all
posterior samples with $N_p=2$ (thus with a total of three Keplerians in the model, accounting for the \OGP) and are therefore available from a single run of
the algorithm. These limits do not rely on injection and recovery of signals
(sinusoidal or otherwise), do not require the subtraction of a best-fit
solution, and marginalize over other orbital parameters, such as eccentricity. The
compatibility limits for the three targets are shown in Fig.~\ref{fig:detlim}.
The adjusted HARPS jitter is larger for \targetII{} than for the other two
targets (see Table~\ref{tab:final_params}), due to a larger scatter in the residuals for this star.
This results in slightly worse detection limits for this source (see Fig.~\ref{fig:detlim}).
For \targetIII{}, the compatibility limits shown in Fig.~\ref{fig:detlim} include PFS RVs in addition to CORALIE and HARPS data.
However, excluding PFS data does not significantly change the detection limits for orbital periods shorter than a year (see Appendix~\ref{sec:compat}).

All three targets were observed by TESS.
We checked whether the inner planets detected for each star  transit in the available TESS data.
This analysis is presented in Appendix~\ref{sec:transitfind}.
We conclude that \targetI{} and \targetII{} are not transiting,
while the known transit of \targetIII{} is clearly recovered \citep[see][]{teske2020}.

To constrain the inclinations of the \OGP
and thereby their true masses, we jointly fit
RV and astrometric data from the Hipparcos-Gaia Catalog of Accelerations \citep[HGCA;][]{Brandt2021}
using \texttt{orvara} \citep{Brandt_etal2021}.
The HGCA provides the $\chi^2$ value for each star,
which should follow a $\chi^2$ distribution with 2 degrees of freedom if the star lacks acceleration.
The $\chi^2$ is 0.709 (p-value of $70\%$) for \targetI{},
64.9 (p-value of $8\times 10^{-15}$) for \targetII{}, and
1.12 (p-value of $57\%$) for \targetIII{}.
Thus, we detect a highly significant acceleration for \targetII{},
and nonsignificant accelerations for \targetI{} and \targetIII{}.
We used a customized version of
\texttt{orvara}\footnote{Available at: \url{https://gitlab.unige.ch/Damien.Segransan/orvara}},
which accommodates a larger set of priors, in particular for between-instrument RV offsets.
All priors were set as in Sect.~\ref{sec:kima}, except for RV jitters (log-uniform distribution from 0.1 \ms to 20 \ms),
and for the semi-major axis (log-uniform from 0.0025~AU to 500~AU)
and mass (log-uniform from $10^{-7} M_{\odot}$ to 1 $M_{\odot}$) of the planets,
which were explored instead of the period and RV semi-amplitude.
We included a Gaussian prior on the period of the \ILP
(the maximum likelihood solution from Table~\ref{tab:final_params}), to facilitate the convergence of the fit.
We excluded the binary companion to \targetII{}, HD~196068, from the \texttt{orvara} fit,
because it lies at a separation of $> 16$ arcseconds \citep[$\approx 800$ AU,][]{GaiaCollaboration2023},
too far to be detected in our RVs or in HGCA astrometry of \targetII{}.

We ran \texttt{orvara} with ten temperatures, each with 150 walkers, for 150,000 steps.
As expected, we are unable to constrain the inclination of the three \ILPs.
We also find no inclination constraints for the \OGP of \targetI{} and \targetIII{},
which is also expected given the nonsignificant Hipparcos-Gaia accelerations
for these two stars.

The posteriors of the \texttt{orvara} orbital fit for \targetII{} are shown in Fig.~\ref{fig:HD196067_orvara}.
We find that the true mass of the \OGP of \targetII{} is $m_b = 11.4 \pm 1.6~M_J$.
The inclination distribution is bimodal, with two almost symmetric modes at
$i = 29.7^{+5.0}_{-3.8}$~deg and $i=149.2^{+3.9}_{-5.1}$~deg.
The mode at 29.7~deg appears slightly favored by the data.
We find a semi-major axis of $4.71^{+0.11}_{-0.10}$~AU.
These results update those presented in \citet{Li2021},
providing finer constraints on the orbital parameters of the \OGP (a factor of two on the semi-major axis)
thanks to our additional RV data.

\section{Discussion}
\label{sec:discussion}

This article presents the three inner planet detections obtained by
monitoring a sample of 26 stars with known outer giant planets using HARPS. A
dedicated statistical study including these 26 stars and additional well-monitored stars with and without outer giant planets will be presented in a
subsequent paper. We note here that two of these inner planets lie in the
SE range (3--10~$m_\oplus$), and one is in the NE
range (10--30~$m_\oplus$), slightly above the SE-NE limit.

The estimated
occurrences of SE and NE below 50~d in the general population are $0.11\pm
  0.03$ and $0.08\pm0.02$, respectively \citep{bashi_2020_occurrence}. If the
occurrence rates of SE and NE were the same in our sample as in the general
population (i.e., without impact from the giant planets), we would typically expect to
detect 2.9 SE and 2.1 NE (about five inner planets in total).
From the compatibility
limits of the three targets presented here (see Fig.~\ref{fig:detlim}),
we see that our observing strategy enables the detection of NE below 50~d
with very high completeness.
For SE, completeness is very poor above 10~d,
and we aim to improve this in the long term with additional HARPS and ESPRESSO
measurements (on-going ESO LP).
Since we used the same observing strategy for the 26 targets in our sample,
we can extrapolate, as a first approximation, that similar detection limits are reached
for the 23 other targets, for which we do not find any \ILP.

We detect fewer inner planets than expected from the general population occurrence rates,
but given the small-number statistics and without a proper statistical analysis correcting for completeness,
we cannot conclude on a slight enhancement or inhibition of \ILP in the presence of \OGP.
However, these results are difficult to reconcile with the claim of \citet{zhu_2018_super,bryan_2019_excess}
that $p(\mathrm{SE}\cup\mathrm{NE} | \mathrm{\OGP})$ is close to 90\%.

Interestingly, the three \ILPs detected in this study all have periods below 6~d.
While this could be due to small-number statistics and reduced detection efficiency at
longer periods (see Fig.~\ref{fig:detlim}),
these three detections are significant, and we should have been able to detect,
albeit with lower confidence,
similar-mass planets up to 20~d .
This observation, if confirmed with a rigorous statistical analysis,
would be in agreement with the findings of \citet{vanzandt_2025_tesskeck}
that \ILPs tend to be found at shorter periods in systems with an \OGP than in the general population.

\FloatBarrier

\section*{Data availability}
The RV data for the three targets discussed in this work, as well as \texttt{kima} posterior samples,
are available on the DACE platform: \url{https://doi.org/10.82180/dace-inner001}.

\begin{acknowledgements}
  Based on observations collected at the European Southern Observatory under ESO programmes
  072.C-0488, 192.C-0852, 0101.C-0275, 0103.C-0785, 1102.C-0923, and 108.22KV.
  The authors acknowledge the financial support of the Swiss National Science
  Foundation (SNSF), supported since May 2022 over the grant 200020\_205010. The
  120~cm EULER telescope and the CORALIE spectrograph were funded by the SNSF
  and the University of Geneva.
  This work has been carried out within the framework of the National Centre of
  Competence in Research PlanetS supported by the SNSF under grants
  51NF40\_182901 and 51NF40\_205606.
  This publication makes use of The Data \& Analysis Center for Exoplanets
  (DACE), which is a facility based at the University of Geneva (CH) dedicated
  to extrasolar planets data visualisation, exchange and analysis. DACE is a
  platform of the Swiss National Centre of Competence in Research (NCCR)
  PlanetS, federating the Swiss expertise in Exoplanet research. The DACE
  platform is available at \href{https://dace.unige.ch}{dace.unige.ch}.
  This research has made use of the SIMBAD database, CDS, Strasbourg
  Astronomical Observatory, France \citep{wenger2000}.
  XD acknowledges the support from the European Research Council (ERC) under the European Union’s Horizon 2020 research and innovation programme (grant agreement SCORE No 851555) and from the Swiss National Science Foundation under the grant SPECTRE (No 200021\_215200).
  NCS is Funded by the European Union (ERC, FIERCE, 101052347). Views and opinions expressed are however those of the author(s) only and do not necessarily reflect those of the European Union or the European Research Council. Neither the European Union nor the granting authority can be held responsible for them. This work was supported by FCT - Fundação para a Ciência e a Tecnologia through national funds by these grants: UIDB/04434/2020 DOI: 10.54499/UIDB/04434/2020, UIDP/04434/2020 DOI: 10.54499/UIDP/04434/2020.
\end{acknowledgements}

\bibliographystyle{aa}
\bibliography{biblio}

\begin{appendix}

  \onecolumn
  \section{Extended posteriors table}
  \label{sec:fulltable}

  \begin{table*}[ht]
  \centering
  \renewcommand{\arraystretch}{1.5}
  \setlength{\tabcolsep}{1pt}
  \caption{Maximum likelihood solution and \texttt{kima} posteriors for \targetI, \targetII, \targetIII.}
  \label{tab:final_params}
  \begin{tabular}{l c c c | c c | c c}
    \hline
    \hline
                                           &                                           & \multicolumn{2}{c|}{\targetI}             & \multicolumn{2}{c|}{\targetII} & \multicolumn{2}{c}{\targetIII}                                                                      \\
    Parameter                              & unit                                      & max-likelihood~\tablefootmark{a}                            & \texttt{kima} posterior~\tablefootmark{b}        & max-likelihood~\tablefootmark{a}                 & \texttt{kima} posterior~\tablefootmark{b} & max-likelihood~\tablefootmark{a} & \texttt{kima} posterior~\tablefootmark{b} \\
    \hline
    \hline
    $v_\mathrm{sys.}$                      & \ms                                       & $573.801 {\,\scriptstyle \pm 0.319 }$
                                           & $574.0 \pm 0.5$
                                           & $-10945.854 {\,\scriptstyle \pm 0.922 }$
                                           & $-10945.0 ^{+1.5} _{-1.6}$
                                           & $19776.047 {\,\scriptstyle \pm 0.825 }$
                                           & $19776.5 ^{+1.6} _{-1.5}$
    \\
    $v_\mathrm{CORALIE98}$                 & \ms                                       & $65.81 {\,\scriptstyle \pm 1.30 }$
                                           & $65.9 \pm 1.4$
                                           & $64.74 {\,\scriptstyle \pm 2.26 }$
                                           & $65.5 ^{+3.0} _{-2.8}$
                                           & $-31.82 {\,\scriptstyle \pm 3.01 }$
                                           & $-33 ^{+4} _{-5}$
    \\
    $v_\mathrm{CORALIE07}$                 & \ms                                       & $70.44 {\,\scriptstyle \pm 1.48 }$
                                           & $70.6 \pm 1.5$
                                           & $61.83 {\,\scriptstyle \pm 1.35 }$
                                           & $62.1 ^{+1.6} _{-1.5}$
                                           & $-36.89 {\,\scriptstyle \pm 1.31 }$
                                           & $-37.6 ^{+1.7} _{-2.9}$
    \\
    $v_\mathrm{CORALIE14}$                 & \ms                                       & $95.196 {\,\scriptstyle \pm 0.634 }$
                                           & $95.2 ^{+0.7} _{-0.6}$
                                           & $88.36 {\,\scriptstyle \pm 1.25 }$
                                           & $88.7 ^{+1.5} _{-1.4}$
                                           & $-17.26 {\,\scriptstyle \pm 1.05 }$
                                           & $-17.6 ^{+1.3} _{-1.5}$
    \\
    $v_\mathrm{HARPS03}$                   & \ms                                       & $-12.112 {\,\scriptstyle \pm 0.525 }$
                                           & $-11.9 \pm 0.7$
                                           & --                                        & --
                                           & --                                        & --                                                                                                                                                                               \\
    $v_\mathrm{PFS1}$~\tablefootmark{c}    & \ms                                       & --                                        & --
                                           & --                                        & --
                                           & $-160.29 {\,\scriptstyle \pm 1.08 }$
                                           & $-160.7 ^{+1.7} _{-2.2}$
    \\
    $v_\mathrm{PFS2}$~\tablefootmark{c}    & \ms                                       & --                                        & --
                                           & --                                        & --
                                           & $-162.46 {\,\scriptstyle \pm 0.832 }$
                                           & $-163.3 ^{+1.6} _{-2.1}$
    \\
    $\sigma_\mathrm{CORALIE98}$            & \ms                                       & $6.46 {\,\scriptstyle \pm 1.10 }$
                                           & $6.8 ^{+1.4} _{-1.2}$
                                           & $6.44 {\,\scriptstyle \pm 1.85 }$
                                           & $9.2 ^{+2.6} _{-2.9}$
                                           & $7.32 {\,\scriptstyle \pm 2.20 }$
                                           & $8.3 ^{+3.1} _{-2.4}$
    \\
    $\sigma_\mathrm{CORALIE07}$            & \ms                                       & $5.28 {\,\scriptstyle \pm 1.21 }$
                                           & $5.4 ^{+1.6} _{-1.3}$
                                           & $8.91 {\,\scriptstyle \pm 1.01 }$
                                           & $8.7 ^{+1.1} _{-1.0}$
                                           & $7.067 {\,\scriptstyle \pm 0.914 }$
                                           & $7.1 ^{+1.1} _{-0.9}$
    \\
    $\sigma_\mathrm{CORALIE14}$            & \ms                                       & $3.019 {\,\scriptstyle \pm 0.651 }$
                                           & $3.0 ^{+0.7} _{-0.8}$
                                           & $4.86 {\,\scriptstyle \pm 1.02 }$
                                           & $4.8 ^{+1.3} _{-1.1}$
                                           & $5.287 {\,\scriptstyle \pm 0.826 }$
                                           & $5.1 ^{+1.0} _{-0.9}$
    \\
    $\sigma_\mathrm{HARPS03}$              & \ms                                       & $1.684 {\,\scriptstyle \pm 0.266 }$
                                           & $1.9 ^{+0.4} _{-0.3}$
                                           & --                                        & --
                                           & --                                        & --                                                                                                                                                                               \\
    $\sigma_\mathrm{HARPS15}$              & \ms                                       & $1.253 {\,\scriptstyle \pm 0.174 }$
                                           & $1.46 ^{+0.27} _{-0.26}$
                                           & $2.381 {\,\scriptstyle \pm 0.215 }$
                                           & $2.48 ^{+0.33} _{-0.25}$
                                           & $1.435 {\,\scriptstyle \pm 0.163 }$
                                           & $1.54 ^{+0.21} _{-0.18}$
    \\
    $\sigma_\mathrm{PFS1}$                 & \ms                                       & --                                        & --
                                           & --                                        & --
                                           & $4.044 {\,\scriptstyle \pm 0.522 }$
                                           & $4.1 ^{+0.7} _{-0.5}$
    \\
    $\sigma_\mathrm{PFS2}$                 & \ms                                       & --                                        & --
                                           & --                                        & --
                                           & $1.590 {\,\scriptstyle \pm 0.408 }$
                                           & $1.8 ^{+0.6} _{-0.4}$
    \\
    \hline
    $P_\mathrm{b}$                         & d                                         & $735.740 {\,\scriptstyle \pm 0.220 }$
                                           & $735.7 ^{+0.4} _{-0.4}$
                                           & $3413.7 {\,\scriptstyle \pm 10.5 }$
                                           & $3397 ^{+17} _{-14}$
                                           & $1609.5 {\,\scriptstyle \pm 19.4 }$
                                           & $1616 ^{+46} _{-28}$
    \\
    $U_\mathrm{0,\,b}${\tiny$=M_\mathrm{0,\,b}+\omega_\mathrm{b}$}\tablefootmark{d}               & deg                                       & $107.069 {\,\scriptstyle \pm 0.563 }$
                                           & $107.0 \pm 0.9$
                                           & $-50.31 {\,\scriptstyle \pm 1.45 }$
                                           & $-53.0 ^{+2.6} _{-2.2}$
                                           & $-70.55 {\,\scriptstyle \pm 8.07 }$
                                           & $-66 ^{+16} _{-12}$
    \\
    $K_\mathrm{b}$                         & \ms                                       & $54.386 {\,\scriptstyle \pm 0.248 }$
                                           & $54.4 ^{+0.3} _{-0.4}$
                                           & $82.57 {\,\scriptstyle \pm 1.91 }$
                                           & $80 ^{+3} _{-4}$
                                           & $7.653 {\,\scriptstyle \pm 0.539 }$
                                           & $7.8 ^{+0.8} _{-0.6}$
    \\
    $e_\mathrm{b}$                         &                                           & $0.07348 {\,\scriptstyle \pm 0.00651 }$
                                           & $0.072 ^{+0.012} _{-0.011}$
                                           & $0.56292 {\,\scriptstyle \pm 0.00703 }$
                                           & $0.561 ^{+0.011} _{-0.012}$
                                           & $0.0389 {\,\scriptstyle \pm 0.0677 }$
                                           & $0.07 ^{+0.09} _{-0.05}$
    \\
    $\omega_\mathrm{b}$\tablefootmark{e}                    & deg                                       & $33.20 {\,\scriptstyle \pm 4.92 }$
                                           & $29 ^{+15} _{-7}$
                                           & $142.835 {\,\scriptstyle \pm 0.966 }$
                                           & $142.0 ^{+1.5} _{-1.8}$
                                           & $201.4 {\,\scriptstyle \pm 70.4 }$
                                           & $37 ^{+100} _{-180}$
    \\
    $m_b\sin i_b$\tablefootmark{f}                          & $M_J$                                     & $2.66 {\,\scriptstyle \pm 0.05 }$
                                           & $2.66 \pm 0.05$
                                           & $5.58 {\,\scriptstyle \pm 0.44 }$
                                           & $5.42 ^{+0.20} _{-0.26}$
                                           & $0.46 {\,\scriptstyle \pm 0.04 }$
                                           & $0.49 ^{+0.05} _{-0.04}$
    \\
    \hline
    $P_\mathrm{c}$                         & d                                         & $5.748913 {\,\scriptstyle \pm 0.000204 }$
                                           & $5.74894 ^{+0.00022} _{-0.00021}$
                                           & $4.601088 {\,\scriptstyle \pm 0.000192 }$
                                           & $4.60111 \pm 0.00024$
                                           & $3.984436 {\,\scriptstyle \pm 0.000217 }$
                                           & $3.98450 ^{+0.00023} _{-0.00022}$
    \\
    $U_\mathrm{0,\,c}${\tiny$=M_\mathrm{0,\,c}+\omega_\mathrm{c}$}\tablefootmark{d}               & deg                                       & $155.61 {\,\scriptstyle \pm 9.54 }$
                                           & $156 \pm 10$
                                           & $-3.3 {\,\scriptstyle \pm 14.4 }$
                                           & $-1 ^{+18} _{-17}$
                                           & $-17.0 {\,\scriptstyle \pm 22.3 }$
                                           & $-10 ^{+24} _{-22}$
    \\
    $K_\mathrm{c}$                         & \ms                                       & $2.702 {\,\scriptstyle \pm 0.252 }$
                                           & $2.65 ^{+0.26} _{-0.27}$
                                           & $3.762 {\,\scriptstyle \pm 0.396 }$
                                           & $3.6 \pm 0.4$
                                           & $2.866 {\,\scriptstyle \pm 0.253 }$
                                           & $2.82 ^{+0.25} _{-0.26}$
    \\
    $e_\mathrm{c}$                         &                                           & $0.097 {\,\scriptstyle \pm 0.124 }$
                                           & $0.07 ^{+0.09} _{-0.05}$
                                           & $0.235 {\,\scriptstyle \pm 0.118 }$
                                           & $0.12 ^{+0.12} _{-0.09}$
                                           & $0.0619 {\,\scriptstyle \pm 0.0817 }$
                                           & $0.05 ^{+0.07} _{-0.04}$
    \\
    $\omega_\mathrm{c}$\tablefootmark{e}                    & deg                                       & $283.4 {\,\scriptstyle \pm 47.2 }$
                                           & $247 ^{+64} _{-166}$
                                           & $295.5 {\,\scriptstyle \pm 23.8 }$
                                           & $278 ^{+41} _{-104}$
                                           & $144.6 {\,\scriptstyle \pm 91.4 }$
                                           & $170 ^{+108} _{-94}$
    \\
    $m_c\sin i_c$\tablefootmark{f}                          & $M_\oplus$                                & $8.26 {\,\scriptstyle \pm 0.79 }$
                                           & $8.1 \pm 0.8$
                                           & $10.4 {\,\scriptstyle \pm 1.39 }$
                                           & $10.1 \pm 1.2$
                                           & $7.45 {\,\scriptstyle \pm 0.7 }$
                                           & $7.7 \pm 0.7$
    \\

    \hline
    \hline
    \multicolumn{2}{r}{reference epoch}    & \multicolumn{2}{c|}{2\,455\,000~BJD}
                                           & \multicolumn{2}{c|}{2\,455\,000~BJD}
                                           & \multicolumn{2}{c}{2\,455\,000~BJD}                                                                                                                                                                                          \\
    \multicolumn{2}{r}{$N$}                & \multicolumn{2}{c|}{199}
                                           & \multicolumn{2}{c|}{187}
                                           & \multicolumn{2}{c}{246}                                                                                                                                                                                                      \\
    \multicolumn{2}{r}{$\log L_{\rm max}$} & \multicolumn{2}{c|}{ -537.8944
    }
                                           & \multicolumn{2}{c|}{ -569.7350
    }
                                           & \multicolumn{2}{c}{ -690.3457
    }                                                                                                                                                                                                                                                                     \\
    \multicolumn{2}{r}{$\log Z$}           & \multicolumn{2}{c|}{-625.97}
                                           & \multicolumn{2}{c|}{-644.7}
                                           & \multicolumn{2}{c}{-767.46}                                                                                                                                                                                                  \\
    \multicolumn{2}{r}{$\mathcal{I}$}      & \multicolumn{2}{c|}{77.19}
                                           & \multicolumn{2}{c|}{67.56}
                                           & \multicolumn{2}{c}{70.37}                                                                                                                                                                                                    \\
  \end{tabular}
  \tablefoot{\tablefoottext{a}{The provided values are the maximum-likelihood estimates, and the errorbars are the 1-$\sigma$ uncertainties estimated from the second derivatives of the log-likelihood.}
  \tablefoottext{b}{The provided values are the posterior medians and the errorbars correspond to the  16\% and and 84\% quantiles.}
  \tablefoottext{c}{
      An arbitrary offset of 19619.947
      m/s (the systemic RV of \targetIII{} measured by Gaia) was added to the PFS RVs, since the data published by \citet{teske2020} were already mean-subtracted.}
  \tablefoottext{d}{Mean argument of latitude at the reference epoch ($U_0=M_0 + \omega=\lambda_0-\Omega$, where $M_0$, $\lambda_0$, $\omega$, and $\Omega$ are the mean anomaly, mean longitude, argument of periastron, and longitude of the ascending node, respectively).
  }
  \tablefoottext{e}{Argument of periastron of the stellar orbit.
  }
  \tablefoottext{f}{The uncertainties on the planetary masses take into account the stellar mass formal uncertainties, but not stellar models systematic errors (see Table~\ref{tab:stellar_parameters}).
  }}
  \renewcommand{\arraystretch}{1}
\end{table*}

  \FloatBarrier

  \twocolumn
  \section{Compatibility limits without PFS data}
  \label{sec:compat}

  The compatibility limits shown in Fig.~\ref{fig:detlim} (\textit{right}) for \targetIII{} include PFS radial velocities in addition to CORALIE and HARPS data.
  This is useful to assess whether additional planets could still hide in this system.
  However, in order to assess the sensitivity of our observing programme to detect small planets, we additionally produced a compatibility map using only CORALIE and HARPS data, which is the typical case for most targets in our programme.

  \begin{figure}[ht]
    \centering
    \includegraphics[width=0.75\linewidth]{figures/HD86226_detlim.pdf}\\
    \includegraphics[width=0.75\linewidth]{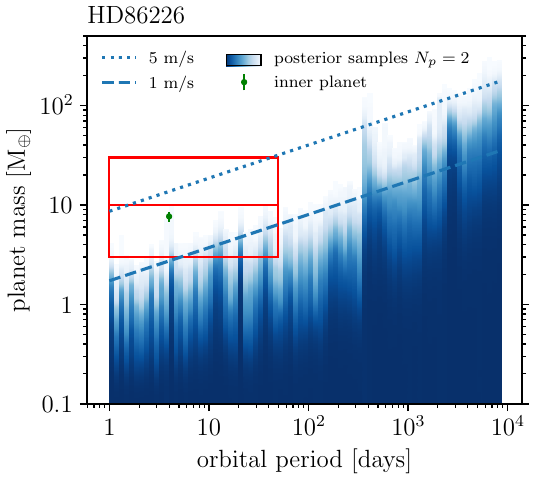}
    \caption{Compatibility limits of \targetIII{} including (\textit{top},
      see also Fig.~\ref{fig:detlim})
      or not (\textit{bottom}) PFS RVs in the analysis.}
    \label{fig:detlim_HD86226}
  \end{figure}

  Both compatibility maps are shown in Fig.~\ref{fig:detlim_HD86226} for comparison.
  Adding the PFS data improves the long-period planets detectability but only marginally the short periods planets detectability.
  Short periods are well covered by HARPS measurements which have a significantly better precision than PFS measurements (see Table~\ref{tab:observations}).
  Thus the addition of PFS data does not improve much the SNR at short period.
  For longer periods, including PFS measurements allows to have a longer baseline covered with high-precision RV instruments.

  \FloatBarrier

  \section{Inner planets transit-search with TESS}
  \label{sec:transitfind}

  In this appendix we present the search for potential transits of the inner planets detected for each star.
  For this analysis, we fix the stellar parameters at their median value reported in Table \ref{tab:stellar_parameters}.
  For each star, the raw PDCSAP flux was downloaded using the {\ttfamily lightkurve}\footnote{\url{https://docs.lightkurve.org/}} package. We obtained sectors 3, 4, 30 and 31 for \targetI{}, sectors 13, 27, 66 and 67 for \targetII{}, and sectors 9, 35 and 62 for \targetIII{}. Then, we split the time series into half-sectors in order to model the noise on each independently.
  We detrend the data with a B-spline using the \texttt{linmarg}\footnote{\url{https://gitlab.unige.ch/delisle/linmarg}} Python package \citep[see][]{leleu_2023_removing}. The spacing of the nodes of the B-spline is chosen to be 3 times the transit duration of a planet on a circular orbit at the period of the RV detection and with a null impact parameter.

  We first performed a BLS search around the period found in RV with a 3$\sigma_{P_c}$ window in period ($\sigma$ is chosen as the max of the two errors reported in Table \ref{tab:final_params}) around the median $\bar{P_c}$, to see if there is an obvious transit in the data. We found \targetIII{}~c with properties similar to the one reported in \cite{teske2020}, but no signal with SNR more than 4 in this pre-detrended dataset for \targetI{}~c and \targetII{}~c.

  \begin{figure}[ht]
    \begin{center}
      \includegraphics[width=0.8\linewidth]{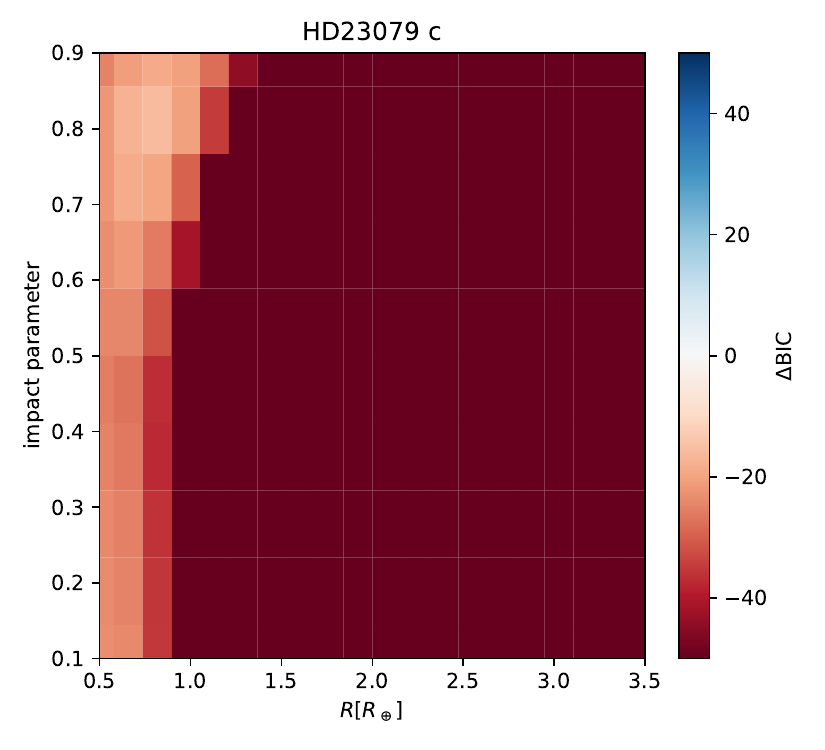}
      \caption{\label{fig:tf_HD23079c} Model comparison map for \targetI{}~c, for radius ranging from 0.5 to 3.5 $R_{\oplus}$ and impact parameters ranging from 0.1 to 0.9. The colors indicate the value of the $\Delta BIC$, red indicating that the null model is favoured, while blue indicates that the planet model is favoured. a value of $|\Delta BIC|>20$ is usually considered to be a compelling evidence for the model. }
    \end{center}
  \end{figure}

  \begin{figure}[ht]
    \begin{center}
      \includegraphics[width=0.8\linewidth]{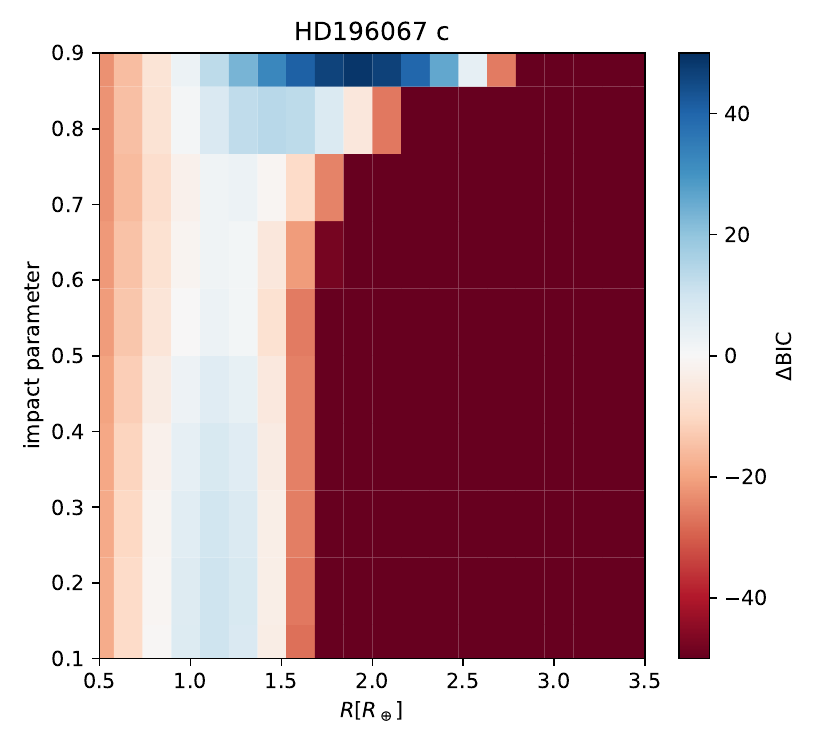}
      \caption{\label{fig:tf_HD196067c} Model comparison map for \targetII{}~c, see Fig. \ref{fig:tf_HD23079c} the text for more detail.}
    \end{center}
  \end{figure}

  \begin{figure}[ht]
    \begin{center}
      \includegraphics[width=0.8\linewidth]{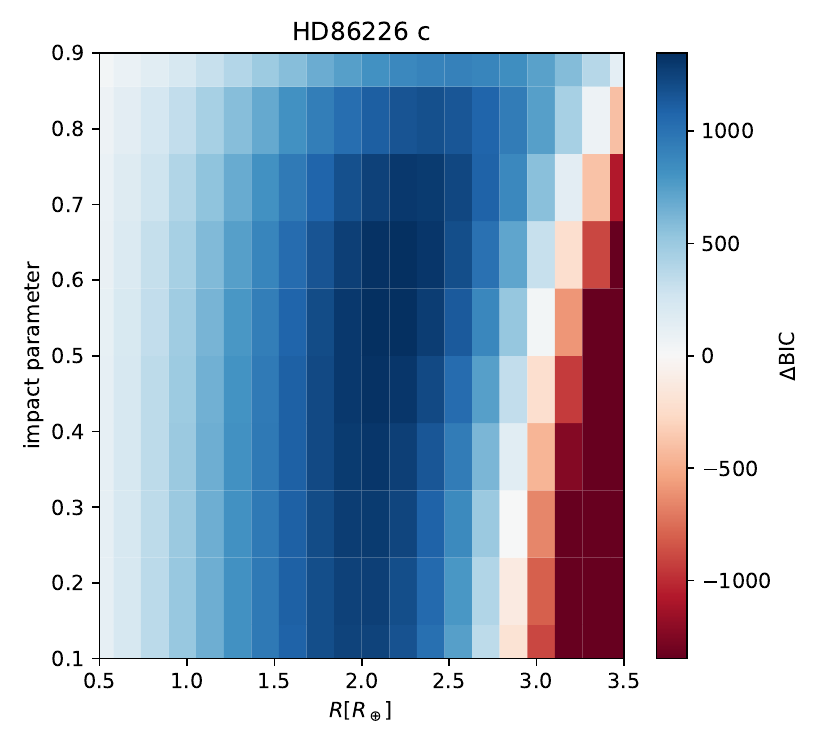}
      \caption{\label{fig:tf_HD86226c} Model comparison map for \targetIII{}~c, see Fig. \ref{fig:tf_HD23079c} the text for more detail.}
    \end{center}
  \end{figure}

  We then computed a model comparison map for each planet with a simultaneous noise and planet modelling. To do so, we first identify potentially problematic parts of the light curves using the pre-detrended data.
  We start by flagging the data points that are at a larger separation from the mean value than 4 times the standard deviation. We then binned the data with a 30~min cadence. We then checked, for each half-sector, if the standard deviation of the flux was 2.5 larger than the minimum standard deviation of all the binned half-sector for that star. This notably lead to the removal of sector 66 for \targetII{} which had a much larger scatter than the others. All other half-sectors were kept.
  We then identified all time stamps that were removed in the detrended light curves, and removed it from the raw data. Restarting from the masked raw data, we then estimated the likelihood of the null model $\mathcal{L}_0$, only applying the detrending. Then, for a grid of planetary radii $R_c$ and impact parameters $b_c$, we explored a grid of orbits ranging the $3\sigma$ interval both in period and $\lambda_0$ for each inner planet.
  The step in $\lambda_{0}$ is set such that the time of transit around the reference epoch varies by 10~min steps.
  The step in period is set such as the last transit varies by steps of 15 min ($\delta P = 6\sigma_{P_c}(t_{max}-t_{min})/\bar{P_c})/96$.
  Then, for each couple ($R_c$, $b_c$), we save the highest likelihood $\mathcal{L}_p$ in the grid ($t_{0,c}$, $P_c$), with a simultaneous modelling of the transits of the planet and the noise model (the aforementioned B-splines).
  We then compute the $\Delta BIC = 2 (\mathcal{L}_p - \mathcal{L}_0) - 4 \log (N_{bins}) $.

  \FloatBarrier

  Figures \ref{fig:tf_HD23079c}, \ref{fig:tf_HD196067c} and \ref{fig:tf_HD86226c} show the model comparison maps for \targetI{}~c, \targetII{}~c and \targetIII{}~c, respectively. For \targetI{}~c and \targetIII{}~c, the interpretation of the maps are straightforward: existing data exclude a planetary transit for \targetI{}~c, while the known $\sim 2 R_{\oplus}$ planet is favoured for \targetIII{}~c \citep{teske2020}. For \targetII{}~c, we notice a $\Delta BIC \sim 48$ for $R_c\sim 1.9 R_{\oplus}$ and an impact parameter of 0.9. We investigated this solution by estimating $\Delta BIC$ per half-sector instead of the whole dataset. It turns out that $\Delta BIC$ is only positive for the first half of sector 13, and negative for the 5 other half-sectors. Upon visual inspection of the lightcurve, a single dip of the flux over a timescale of $\sim 1h$ seems to be responsible for the positive value of the $\Delta BIC$ in the first half of sector 13. We therefore conclude that the planet is not transiting.

\end{appendix}

\end{document}